\documentclass[twocolumn]{aastex631}

\usepackage[colorinlistoftodos]{todonotes}
\usepackage{amssymb}
\usepackage{comment}
\usepackage{multirow,acronym}
\usepackage{natbib}
\usepackage{makecell}
\usepackage{booktabs}
\usepackage{amsmath}
\usepackage{subfigure}
\usepackage{color}
\usepackage{xcolor}

\usepackage{hyperref}
\usepackage[T1]{fontenc}
\usepackage{graphicx}
\usepackage{bm}
\usepackage{CJK}
\hypersetup{colorlinks=true, citecolor=blue, 
  linkcolor=cyan, urlcolor=magenta}

\usepackage{aas_macros}
\usepackage{float}
\usepackage{amsmath,amssymb}
\usepackage{natbib} 
\usepackage{graphicx} 
\usepackage{color}
\usepackage{verbatim}
\usepackage{rotating}

\newcommand{\proptosim}{\mathrel{\vcenter{
 \offinterlineskip\halign{\hfil$##$\cr
 \propto\cr\noalign{\kern2pt}\sim\cr\noalign{\kern-2pt}}}}}
\renewcommand{\b}[1]{\boldsymbol{#1}}
\renewcommand{\i}{\ensuremath{\rm i}}
\newcommand{\unit}[1]{{\rm\, #1}}

\newcommand{\response}[1]{{#1}}

\newcommand{\mean}[1]{\langle #1\rangle}

\renewcommand{\max}{\mathrm{max}} 
\newcommand{\au}{\mathrm{AU}}

\newcommand{\m}{\unit{m}}
\newcommand{\g}{\unit{g}}
\newcommand{\G}{\unit{G}}
\newcommand{\K}{\unit{K}} 
\newcommand{\km}{\unit{km}}

\newcommand{\s}{\mathrm{s}}
\newcommand{\yr}{\mathrm{yr}}

\newcommand{\ang}{\ensuremath{\mathrm{\AA}}}

\newcommand{\B}{\mathbf{B}}     
\newcommand{\E}{\mathbf{E}}     
\newcommand{\J}{\mathbf{J}}     
\newcommand{\F}{\mathbf{F}}     
\renewcommand{\v}{\mathbf{v}}   

\renewcommand{\d}{\mathrm{d}}
\newcommand{\e}{\mathrm{e}}

\renewcommand{\ion}[2]
{{\rm#1}\;\textsc{#2}}
\newcommand{\p}{\mathrm{p}}     

\newcommand{\crit}{\mathrm{crit}}

\renewcommand{\O}{\mathrm{O}}     
\newcommand{\A}{\mathrm{A}}     
\renewcommand{\H}{\mathrm{H}}     
\renewcommand{\P}{\mathrm{P}}     

\renewcommand{\i}{\ensuremath{\mathrm{i}}}
\newcommand{\figdir}{.}

\begin{document}

\title{Non-ideal Magnetohydrodynamic Instabilities in
  Protoplanetary Disks: \\ Vertical Modes and Reflection
  Asymmetry}

\author[0000-0002-6540-7042]{Lile Wang}
\affil{The Kavli Institute for Astronomy and Astrophysics,
  Peking University, Beijing 100871, China}
\affil{Department of Astronomy, School of Physics, Peking
  University, Beijing 100871, China}

\author[0000-0003-2895-4968]{Sheng Xu}
\affil{The Kavli Institute for Astronomy and Astrophysics,
  Peking University, Beijing 100871, China}
\affil{Department of Astronomy, School of Physics, Peking
  University, Beijing 100871, China}

\author[0000-0003-1264-2222]{Zhenyu Wang}
\affil{Institute of Plasma Physics, Chinese Academy of
  Sciences, Hefei 230031, China}
\affil{Department of Astrophysical Sciences, Princeton
  University, Princeton NJ 08540, USA} 

\author[0000-0001-8060-1321]{Min Fang}
\affil{Purple Mountain Observatory,
  Chinese Academy of Sciences, 10 Yuanhua Road, Nanjing
  210023, China}
\affil{University of Science and Technology
  of China, Hefei 230026, China}

\author[0000-0002-6710-7748]{Jeremy Goodman}
\affil{Department of Astrophysical Sciences, Princeton
  University, Princeton NJ 08540, USA}

\correspondingauthor{Lile Wang}
\email{lilew@pku.edu.cn}
\correspondingauthor{Zhenyu Wang}
\email{zhenyu.wang@ipp.ac.cn}

\begin{abstract}
  Magnetized disk winds and wind-driven accretion are an
  essential and intensively studied dispersion mechanism of
  protoplanetary disks. However, the stability of these
  mechanisms has yet to be adequately examined. This paper
  employs semi-analytic linear perturbation theories under
  non-ideal magnetohydrodynamics, focusing on disk models
  whose magnetic diffusivities vary by a few orders of
  magnitude from the disk midplane to its surface.  Linear
  modes are distinguished by their symmetry with respect to
  the midplane. These modes have qualitatively different
  growth rates: symmetric modes almost always decay, while
  at least one anti-symmetric mode always has a positive
  growth rate. This growth rate decreases faster than the
  Keplerian angular velocity with cylindrical radius $R$ in
  the disk and scales steeper than $R^{-5/2}$ in the
  fiducial disk model. The growth of anti-symmetric modes
  breaks the reflection symmetry across the disk equatorial
  plane, and may occur even in the absence of the Hall
  effect. In the disk regions where fully developed
  anti-symmetric modes occur, accretion flows appear only on
  one side of the disk, while disk winds occur only on the
  other.  This may explain the asymmetry of some observed
  protoplanetary disk outflows.
\end{abstract}


\section{Introduction}
\label{sec:intro}

Protoplanetary disks (PPDs), which serve as the cradles for
planet formation, undergo a lifespan of approximately
$10^6-10^7~\yr$. They follow three distinct processes to
disperse, namely: (1) creation of planets, (2) accretion
onto the central protostar, and (3) outflowing in winds. The
latter two processes directly compete with the first,
impeding the time and mass available to create
planets. Compared to photoevaporative winds that are mostly
unable to drive accretion \citep{2012MNRAS.422.1880O,
  2017ApJ...847...11W}, magnetized winds of PPD exert torque
on the disks. Magnetic fields create the linkage between
disk winds and accretions and have recently been identified
as a crucial factor in PPDs \citep{2013ApJ...769...76B,
  2013ApJ...772...96B, 2016ApJ...818..152B,
  2017ApJ...845...75B, 2019ApJ...874...90W}. In contrast,
other prospective mechanisms have inadequate efficiency in
viscous turbulent accretion via the magnetorotational
instability (MRI; e.g. \citealt{1998RvMP...70....1B}, and
the situation affected by the Ohmic resistivity,
e.g. \citealt{1999ApJ...515..776S}) or any other
hydrodynamic instabilities under PPD conditions (e.g.,
\citealt{2013ApJ...769...76B, 2013ApJ...772...96B,
  2013ApJ...764...66S, 2013ApJ...775...73S}; see
\citealt{2014prpl.conf..411T} for a review).

In recent years, several studies have been conducted to
investigate the magnetized wind-driven accretion process in
protoplanetary disks. In one such study
\citep{2017ApJ...845...75B}, global simulations were
performed using 2.5-dimensional axisymmetric full MHD
simulations with non-ideal MHD effects. Magnetic
diffusivities were evaluated through a pre-calculated
interpolation table, and thermodynamics were calculated via
a simple relaxation-time recipe that accounted for
temperature dependence on spatial
location. \citet{2019ApJ...874...90W} adopted
non-equilibrium thermochemical networks that co-evolved with
non-ideal MHD to consistently determine all magnetic
diffusivities and thermodynamic properties in real
time. Building on these works, \citet{2020ApJ...904L..27N}
performed calculations to predict observational evidence of
magnetized disk winds and the resulting predictions were
subsequently confirmed in recent
observations. High-resolution observations of the
[\ion{O}{i}] $6300~\ang$ emission line were conducted by
\citet{2023NatAs...7..905F}, who confirmed that magnetized
disk winds, rather than photoevaporative winds, were
essential in explaining the spatially resolved emission line
features. These studies highlight the importance of
non-ideal MHD effects and the role of magnetic fields in
understanding the complex dynamics of protoplanetary disks
and their associated winds. There have been, nevertheless,
no direct measurements of disk magnetic fields. Attempts up
to now only yield the upper limits of field strengths in a
handful of disks (e.g., TW Hya, \citealt
{2019A&A...624L...7V}; AS 209, \citealt
{2021ApJ...908..141H}).

Reflection symmetry over the equatorial plane has been
commonly assumed in research on magnetized winds in
protoplanetary disks, mainly concentrating on their general
features, such as launching mechanisms, observables, and
kinematics. Until recently, asymmetries in magnetic winds
have been largely overlooked, with only a few observations
indicating their existence, such as in the case of HH 30
\citep{2004ApJ...602..860W}. However, with advancements in
observational techniques, more observations are expected to
emerge in various wavelength bands. Understanding these
asymmetries is crucial for future observations that depict
detailed characteristics of magnetized disk
winds. \citet{2010MNRAS.406..848L} discovered the growth of
modes with different types of reflection symmetry properties
under the ideal MHD limit.  \citet{2017ApJ...836...46B}
discussed the Hall effect and attributed the breaking of
reflection symmetry to it.  \citet{2017A&A...600A..75B} also
noticed the asymmetric accretion and wind flows in global
simulations with all non-ideal MHD effects included, and
attributes this effect semi-quantitatively to the expulsion
of electric current sheet from the equatorial
plane. Nevertheless, some recent simulation efforts
discovered that the asymmetry may still emerge when the Hall
effect is intentionally turned off while growth rates seem
to continue.  For example, \citet{2015ApJ...801...84G}
noticed the asymmetric pattern of accretion and wind
launching in absence of the Hall effect, with Ohmic
diffusivity and ambipolar diffusion only. With local
shearing box simulations, \citet{2020MNRAS.498..750L}
confirmed the growth of such anti-symmetric modes with Ohmic
diffusivity only.  Interestingly,
\citet{2024MNRAS.530.5131S} reported symmetric accretion
pattern with the Hall effect involved.  This paper will
delve into the symmetry of wind-driven accreting
protoplanetary disk systems, paying attention to their
physical origins involving magnetohydrodynamics and
thermochemistry and their potential observables.

This paper is structured as follows. \S\ref{sec:method}
offers a detailed description of the physical models and
equations used to analyze the given system. Simplified
versions of the model are applied to idealized systems for
qualitative discussions in \S\ref{sec:simp-modes}. The
mathematical models are subsequently applied to the
numerical model cited in \citet{2020ApJ...904L..27N} to
study its asymmetric instabilities, elaborated in
\S\ref{sec:fid-model}. Furthermore, in
\S\ref{sec:var-models}, we explore the parameter space to
understand how different physical assumptions impact the
instabilities studied. Finally, \S\ref{sec:summary} provides
discussions on possible generalizations and observable tests
and presents a summary of the principal findings and
conclusions. Some details of analytic derivations are
presented in the Appendices.

\section{Method}
\label{sec:method} 

The mathematical model of this work is based on the physical
picture of wind-driven accretion in PPDs. Unless otherwise
specified, this paper uses cylindrical coordinates
${R, \varphi, z}$ for describing everything associated with
the geometries, where the $z$-axis is aligned with the PPD's
axis of rotation.

\subsection{Non-ideal MHD for wind-driven accretion}
\label{sec:non-ideal-mhd}

This work adopts non-ideal MHD equations to describe the 
accretion and winds of magnetized PPDs, 
\begin{equation}
  \label{eq:method-mhd}
  \begin{split}
  & \partial_t \rho + \nabla \cdot (\rho \v) = 0 , \\
  & \rho\partial_t \v + \rho \v \cdot \nabla \v =
    \frac{\J}{c} \times \B - \rho \nabla \Phi - \nabla p  ,
    \\
  & \nabla \cdot \B = 0  , \  \nabla \times \B = \frac{4
    \pi}{c} \J ,\\
  & \partial_t \B = - c \nabla \times \E  ,\  \nabla \cdot
    \E = 0  . 
  \end{split}
\end{equation}
Here we use $\rho$ for the mass density, $\v$ the gas
velocity ($c$ the vacuum speed of light), $p$ the gas
pressure, $\Phi$ the gravitational potential, $\J$ the
electric current density, and $\B$ and $\E$ the magnetic and
electric fields. For non-ideal MHD systems with finite
conductivities, the $\E$ field is evaluated by the electric
field in the local fluid rest frame $\E'$,
\begin{equation}
  \label{eq:method-non-ideal}
  \begin{split}
    \E' & = \E + \frac{\v}{c} \times \B \\
    & = \frac{4 \pi}{c^2} \left[ \eta_\O \J + \eta_\H
      \frac{\J \times \B}{|\B|} + \eta_\A \frac{\B
      \times (\J \times \B)}{|\B|^2} \right]  ,
  \end{split}
\end{equation}
where $\eta_{\O,\H,\A}$ are used to denote the Ohmic
($\eta_\O$), Hall ($\eta_\H$), and ambipolar ($\eta_\A$)
components of magnetic diffusivities \citep[see also][]
{2007Ap&SS.311...35W, 2016ApJ...819...68X}.

While previous studies like \citet{2019ApJ...874...90W} have
already determined diffusivity profiles through
self-consistent calculations based on non-equilibrium
thermochemistry for magnetized wind-driven PPD simulations,
this paper aims to explore the potential impact of
hypothetical diffusivity profiles on the dynamical stability
of magnetized PPDs. This approach complements previous
studies and contributes to the ongoing efforts to develop
more accurate and comprehensive models of the disks. Using
hypothetical diffusivity profiles allows for the exploration
of a range of possible scenarios and the identification of
patterns or effects that may not have been previously
considered, deepening our understanding of the physics of
PPDs and informing future research efforts.

\subsubsection{Radially local problems}
\label{sec:method-radial-local}

In order to explore the breaking of reflection symmetry, it
is necessary to have steady-state, symmetric solutions as a
foundation for perturbation theories. Axisymmetric models
local in cylindrical radius $R$ are used to discuss
reflection symmetry properly. In most radial ranges of
typical PPDs, the vertical density scale height $h$ is much
smaller than $R$, which means that the derivatives of
physical variables with respect to $R$ are much smaller than
those with respect to $z$. It is assumed that the disk gas
temperature $T$ is only a function of $R$. This assumption
is a reasonable approximation within PPDs where $T$ is
primarily controlled by thermal accommodation with dust
grains, which are in tight thermal equilibrium with the
radiation emanating from the central star
\citep{1997ApJ...490..368C}. It may not hold very well above
the disk surfaces, such as inside magnetized winds
\citep{2019ApJ...874...90W}, but our analysis focuses
primarily on the behaviors below disk surfaces. For more
explicit discussions, we construct isothermal models
dependent only on $z$, with most radial derivatives
$\partial_R$ considered negligible unless otherwise
specified. Because of the axisymmetry, all azimuthal
derivatives ($\partial_\varphi$) vanish. 

To better regularize and generalize the analyses in this
paper, conversions to dimensionless variables are
necessary. For the radially local isothermal models we
discuss, we introduce the following conversions:
\begin{equation}
  \begin{split}
  \label{eq:method-dimless-trans}
  & \zeta \equiv \frac{z}{h},\ \tau \equiv \Omega_\K t,\ 
    \partial_z\rightarrow\frac{\mu_k}{R} \partial_{\zeta},\ 
    \partial_t \rightarrow \Omega_\K \partial_{\tau} ; \\ 
  & \mu_k \equiv \frac{\Omega_\K R}{c_s} , \
    \mu_\varphi \equiv \frac{v_{\varphi}}{c_s} - \mu_k ,\ 
    \mu_R \equiv \frac{v_R}{c_s},\
    \mu_z \equiv \frac{v_z}{c_s};\\
  & \varrho \equiv \frac{\rho}{\rho_0}  , \
    \beta_0 \equiv = \frac{8 \pi c_s^2\rho_0}{B_{z0}^2},\ 
    b_i \equiv \frac{B_i}{B_{z 0}}, \\
  & \varepsilon_i \equiv \frac{c E_i}{c_s B_{z 0}}, \
    j_i \equiv \frac{4 \pi R J_i}{c B_{z 0} \mu_k}, \
    i \in \{ R, \varphi, z \} .
  \end{split}
\end{equation}
Here $\Omega_\K$ is the angular velocity of the Keplerian
orbital motion, $\rho_0$ is the mid-plane mass density,
$B_{z 0}\equiv B_z|_{z=0}$ is the $z$-component of the
magnetic fields at the equatorial plane, and the isothermal
sound speed satisfies $c_s^2 = p/\rho$. The scale height is
adopted as $h=c_s/\Omega_\K$. The Mach number of Keplerian
velocity $\mu_k$ compares the vertical sound crossing
timescales to the orbital timescales. Its value can be
estimated by,
\begin{equation}
  \label{eq:method-muk}
  \begin{split}
  \mu_k \simeq & 29\times \left( \frac{M_*}{M_{\odot}}
  \right)^{1/2} \left( \frac{T}{300~\K} \right)^{-1/2} \\
    & \quad \times \left( \frac{R}{\au} \right)^{- 1 / 2}
  \left( \frac{\langle m_{\rm mol}  \rangle}{2.35 m_p}
      \right)^{1 / 2}  ,
  \end{split}
\end{equation}
in which $M_*$ is the stellar mass, $m_p$ is the proton
mass, and $\mean{m_{\rm mol}}$ is the mean molecular
mass. It can be easily verified that almost all radial
derivatives in the dimensionless forms are multiplied by
$\mu_k^{-1}$, meaning that their effects on the equations
are suppressed by more than one order of magnitude. The most
important exception is $\partial_RE_z$, in which the term
$\partial_R(v_\varphi B_R)$ reflects the radial velocity
shear, and leads to the $(-3b_R/2)$ term in the
$\partial_\tau b_\varphi$ expression without the
$\mu_k^{-1}$ suppression \citep[see also][]
{Wardle+Konigl1993}. One direct consequence of this
approximation is that the solenoidal condition of magnetic
fields
$\nabla\cdot \B = R^{-1}\partial_R(R B_R) + \partial_z B_z =
0$ reduces to $\partial_z B_z = 0$, and hence $b_z = 1$
always holds. Simulations of magnetized PPD winds have
confirmed this assumption with high confidence
\citep[e.g.][]{2017ApJ...845...75B}.

Under such transforms, the MHD differential
eqs.~\eqref{eq:method-mhd} and \eqref{eq:method-non-ideal}
are recast into dimensionless differential equations for
the axisymmetric, radial local system,
\begin{equation}
  \label{eq:method-system-ode}
  \begin{split}
  & \partial_{\tau} \mu_z + \partial_{\zeta} \ln \varrho + \mu_z
    \partial_{\zeta} \mu_z = \frac{2}{\beta_0 \varrho} (j_R
    b_{\varphi} - j_{\varphi} b_R) - \zeta  ,\\
  & \partial_{\tau} \mu_R + \mu_z \partial_{\zeta} \mu_R =
    \frac{2}{\beta_0 \varrho} (j_{\varphi} b_z - j_z
    b_{\varphi}) + 2 \mu_{\varphi} + g_R, \\ 
  & \partial_{\tau} \mu_{\varphi} + \mu_z \partial_{\zeta}
    \mu_{\varphi} = \frac{2}{\beta_0 \varrho} (j_z b_R - j_R
    b_z) - \frac{\mu_R}{2}  ,  \\
  & \partial_{\tau} b_{\varphi} = - \partial_{\zeta}
      \varepsilon_R - \frac{3}{2} b_R  ,\quad \partial_{\tau}
      b_R = \partial_{\zeta} \varepsilon_{\varphi} . \\ 
  \end{split}  
\end{equation}
Here $g_R$ is a correction term for reducing radial
gravitational force at relatively high altitudes (see
Appendix~\ref{sec:apdx-xtra-terms}). Note that such $g_R$
term only directly affects the disk accretion, not the wind
launching regions or the acceleration processes within. The
dimensionless electric current densities $j$ and fluid frame
electric fields $\varepsilon'$ are related to the reduced
fields $b$ and the rest frame electric fields $\varepsilon$
by,
\begin{equation}
  \label{eq:method-system-efield}
  \begin{split}
  & j_R = - \partial_{\zeta} b_{\varphi} ,\
    j_{\varphi} = \partial_{\zeta}   b_R  , \\
  & \varepsilon'_R = \varepsilon_R + (\mu_{\varphi} + \mu_k)
    b_z - \mu_z b_{\varphi}  , \\
  & \varepsilon'_{\varphi} = \varepsilon_{\varphi} + \mu_z
    b_R - \mu_R b_z  ,\ \varepsilon'_z\simeq 0 .
  \end{split} 
\end{equation}
The value of $j_z$ is generally considered tiny, yet in some
conditions, its contribution is not negligible;
Appendix~\ref{sec:apdx-xtra-terms} also estimates the $j_z$
values. Similar to the dimensional case, the dimensionless
$\varepsilon'$ and $j$ are related by the magnetic
diffusivities,
\begin{equation}
  \label{eq:method-system-nonideal}
  \begin{split}
  & \begin{bmatrix}
    \varepsilon'_R\\ \varepsilon_{\varphi}'
    \end{bmatrix} =
    \begin{bmatrix}
    \alpha_{R R} & \alpha_{R \varphi}\\
    \alpha_{\varphi R} & \alpha_{\varphi \varphi}
    \end{bmatrix}
    \begin{bmatrix} j_R\\ j_{\varphi} \end{bmatrix} , \\
  & \alpha_{R R} \equiv \alpha_{\O} + \frac{\alpha_{\A}}{b^2}
  (b_{\varphi}^2 + 1) ,\ \alpha_{R \varphi} \equiv
    \frac{\alpha_{\H}}{b} - \frac{\alpha_{\A}}{b^2} b_R
    b_{\varphi} , \ \\ 
 &  \alpha_{\varphi R} \equiv - \frac{\alpha_{\H}}{b} -
  \frac{\alpha_{\A}}{b^2} b_R b_{\varphi} ,\
    \alpha_{\varphi \varphi} \equiv \alpha_{\O} +
    \frac{\alpha_{\A}}{b^2} (b_R^2 + 1)\ ; \\
  & \alpha_i \equiv \frac{\eta_i}{h c_s} \equiv \frac{2 b^2}
  {\beta_0 \varrho}\Lambda_i^{-1},\ i \in \{ \O, \H, \A\} .
  \end{split}
\end{equation}
Here $b^2\equiv b_R^2+ b_\varphi^2 + 1$, and we also
introduce $\Lambda_i\equiv B^2/(4\pi\rho\eta_i\Omega_\K) $
for the Elsasser numbers of the three components. It is
already known that the $\eta_\O$, $\eta_\H/|\B|$, and
$\eta_\A/|\B|^2$ are mostly independent of $\B$
\citep[e.g.][] {2016ApJ...819...68X}. This fact implies that
we should adopt $\alpha_\O$, $\alpha_\H/b$, and
$\alpha_\A/b^2$ for the diffusivity profiles in practice.

\subsubsection{Steady states}
\label{sec:method-steady}

To solve eqs.~\eqref{eq:method-system-ode} for the steady
states, it is necessary to set $\partial_\tau\rightarrow
0$. Additionally, a few extra approximations can be made to
simplify the problem further. In the accretion layer,
vertical gas motion $\mu_z$ is significantly smaller than
the horizontal motion $\mu_R$ and $\mu_\varphi$.  The
smallness of $\mu_z$ makes the steady-state versions of
eqs.~\eqref{eq:method-system-ode} stiff via the operator
$\mu_z\partial_z$; yet this operator is unimportant for
vertically smooth solutions. Therefore, setting $\mu_z$ to
zero for the system is practical. The transport of magnetic
fluxes mainly affects $\varepsilon_\varphi$, which is
related to radial transport, that is, $\partial_tB_z$. For a
radial local solution, it is safe to assume that
$\varepsilon_\varphi$ is zero, allowing us to concentrate on
the vertical modes. Eq.~\eqref {eq:method-system-ode} are
simplified by these assumptions into the ordinary
differential equations (ODEs),
\begin{equation}
  \label{eq:method-steady-ode} 
  \begin{split}
  & \partial_{\zeta} \ln\varrho = \dfrac{2}{\beta_0
    \varrho} (j_R b_{\varphi} - j_{\varphi} b_R) - \zeta ,\ 
    \partial_{\zeta} \varepsilon_R = - \dfrac{3}{2} b_R ,\\
  & \partial_\zeta b_R = j_\varphi ,\
    \partial_\zeta b_\varphi = -j_R  ,
  \end{split}
\end{equation}
where the dimensionless current densities $j_{R,\varphi}$
and the subsequent velocity components
are determined by solving the combination of eqs.~\eqref
{eq:method-system-efield} and \eqref
{eq:method-system-nonideal} (see also
Appendix~\ref{sec:apdx-current}).

The symmetry required for the steady state solutions leads
to $b_R = b_\varphi = 0$ at $\zeta = 0$, and
$\varrho|_{\zeta = 0} = 1$ by definition. The free parameter
to be determined is
$\varepsilon_{R0} \equiv \varepsilon_R |_{\zeta = 0}$, whose
value is determined by matching physical parameters of wind
solutions constructed with ideal MHD. Given proper profiles
of diffusivities, eqs.~\eqref{eq:method-steady-ode} are
integrated to an altitude at which the diffusivity is
sufficiently low, characterized by a critical dimensionless
Ohmic diffusivity $\alpha_{\O, c}$. In practice, we find
that $\alpha_{\O, c} = 10^{-4}$ appears to be a good cutoff,
around which small variations of $\alpha_{\O c}$ will not
change the results significantly. At this altitude, an ideal
MHD wind solution is generated using the
\citet{2016ApJ...818..152B} scheme, whose Bernoulli
parameter should match the value given by the steady-state
``disk'' solution described here. We refer the readers to
Appendix~\ref{sec:apdx-matching-wind} for more details.

\subsection{Linear perturbations}
\label{sec:linear-perturb}

The stability of steady-state solutions
\S\ref{sec:method-steady} should be analyzed by perturbing
eqs.~\eqref{eq:method-system-ode}. For any relevant
dependent variables $x$, we decompose it into the form
$x \rightarrow x + \delta x$, where the $x$ now stands for
the steady-state solution, and $\delta x$ for the
perturbation. This decomposition is applied to
eqs.~\eqref{eq:method-system-ode},
\eqref{eq:method-system-efield} and
\eqref{eq:method-system-nonideal}. The perturbations of
variables are preserved only up to the first order. Similar
to \S\ref{sec:method-steady}, the perturbations of vertical
velocities also vanish ($\delta\mu_z\rightarrow 0$) as the
focus of our stability analyses is located below the disk
surface.

To analyze the time evolution of the perturbations, we
assume that all perturbation terms have the same time
dependence, $\delta x\propto \e^{\nu\tau}$, where $\nu$ is
the dimensionless growth rate. The sign of $\nu$ determines
whether a perturbative mode is stable ($\nu \leq 0$) or
unstable ($\nu>0$), and the absolute value of $\nu$
indicates how fast the mode grows or decays.  The momentum
equations for the perturbations then read,
\begin{equation}
  \label{eq:method-perturb-mom}
  \nu \delta \mu_R - 2 \delta \mu_{\varphi} = \frac{2 \delta
    j_{\varphi}}{\beta_0 \varrho}  , \ 
  \frac{\delta \mu_R}{2} + \nu \delta \mu_{\varphi} = -
  \frac{2 \delta j_R}{\beta_0 \varrho}  .
\end{equation}
The derivatives of the perturbed velocities are then,
\begin{equation}
  \label{eq:method-perturb-deriv-mom}
  \begin{split}
    & \partial_{\zeta} \delta \mu_R  =  \zeta \delta \mu_R -
      \left[ \frac{2}{\beta_0 \varrho (1 + \nu^2)} \right]
      (2 \partial_{\zeta} \delta j_R - \nu \partial_{\zeta} 
      \delta j_{\varphi}) ,\\ 
    & \partial_{\zeta} \delta \mu_{\varphi} =  \zeta \delta
      \mu_{\varphi} - \left[ \frac{2}{\beta_0 \varrho (1 +
      \nu^2)} \right] \left( \nu \partial_{\zeta} \delta j_R
      + \frac{ \partial_{\zeta} \delta j_{\varphi}}{2}
      \right) . 
  \end{split}
\end{equation}
To further simplify the analyses, we make two extra
assumptions: (1) that $\delta(\ln\varrho)= 0$ (since
$\mu_z \rightarrow 0$ and $\delta\mu_z = 0$) as the linear
perturbations are insufficient to reshape the vertical
density profile, and (2) that the matrix elements
$\alpha_{ij}$ ($i, j\in\{ R, \varphi\}$) in
eqs.~\eqref{eq:method-system-nonideal} are not susceptible
to the perturbations either. Taking the perturbations into
the field equations in eqs.~\eqref{eq:method-system-ode},
\eqref{eq:method-system-efield} and
\eqref{eq:method-system-nonideal}, one obtains the governing
equations of the perturbations, the matrix form,
\begin{equation}
  \label{eq:method-perturb-ode}  
  \partial_\zeta [(A + \xi V_1)\delta b] = V_0 \delta b ,
\end{equation}
where
$\delta b\equiv [\delta b_R, \delta b_{\varphi}, \delta j_R,
\delta j_{\varphi}]^T$ is the vector of perturbation
variables, and the other involved matrices are defined as,
\begin{equation}
  \label{eq:method-perturb-matrix}
  \begin{split}
    & A \equiv 
    \begin{bmatrix}
    &  & \alpha_{\varphi R} - 2 \xi
    & \alpha_{\varphi\varphi}\\ 
    &  & - \alpha_{R R} 
    & - \alpha_{R \varphi} - \xi / 2\\ 
    & - 1 &  & \\
    1 &  &  & 
    \end{bmatrix} ,\\
    & V_0 \equiv
     \begin{bmatrix}
       \nu &  & &\\ 
       3/2 & \nu &  & \\
           &  & 1 & \\
           &  &  & 1
     \end{bmatrix} ,\quad
    V_1 \equiv
     \begin{bmatrix}
    &  & & \nu\\ 
    &  & -\nu & \\
    & 0 &  & \\
    0 &  &  & 
     \end{bmatrix} ,
  \end{split}
\end{equation}
in which $\xi \equiv 2 / [\beta_0 \varrho (1 + \nu^2)]$, and
we have used the approximation
$\partial_\zeta \xi = \zeta\xi$ which holds inside the
concerned regions (below the wind
base).

\begin{figure}
  \centering
  \hspace*{-0.4cm}
  \includegraphics[width=3.5in, keepaspectratio]
  {\figdir/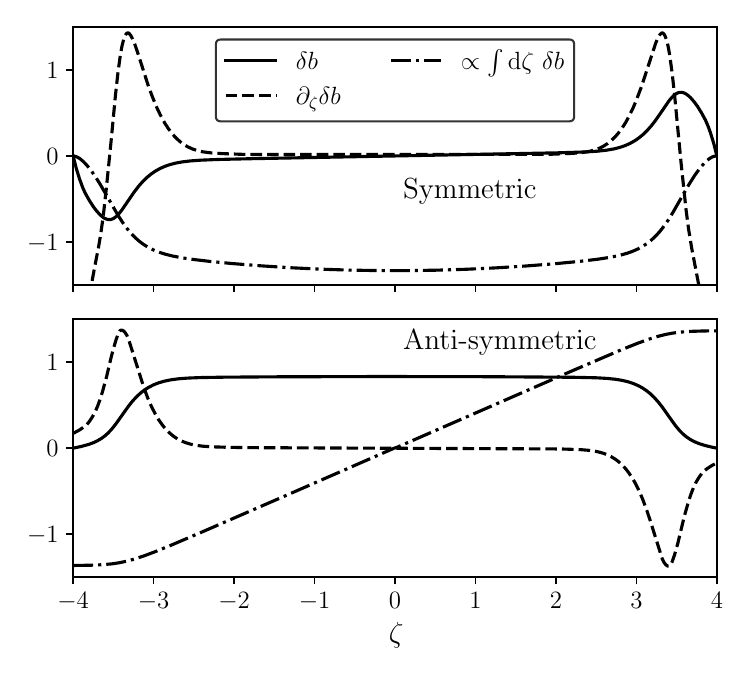}
  \caption{Schematic illustration (not to scale) of
    symmetric (upper panel) and anti-symmetric (lower panel)
    perturbation modes. Note that the symmetries are
    indicated in terms of field line geometries (which
    should refer to $\int\d\zeta\;\delta b$), {\it not} the
    values of $\delta b$ (\S\ref{sec:linear-perturb}).  }
  \label{fig:sym_schematic} 
\end{figure}

Eqs.~\eqref{eq:method-perturb-ode} are integrated with the
eigenvalue $\nu$ from $\zeta = 0$ to the wind base altitude
$\zeta_{\rm wb}$ where $\mu_R$ changes the sign. Two types
of symmetries are possible for $b_{R, \varphi}$:
\begin{itemize}
\item Symmetry: $\delta b_i = 0$ but
  $\partial_{\zeta} \delta b_i \neq 0$ at $\zeta = 0$;
\item Anti-symmetry: $\delta b_i \neq 0$ but
  $\partial_{\zeta} \delta b_i = 0$ at $\zeta =
  0$.
\end{itemize}
Note that, similar to \S\ref{sec:non-ideal-mhd}, the
definitions of ``symmetry'' and ``anti-symmetry'' are for
the field line morphologies, viz. the integral curves of
$\delta b$ (see Figure~\ref{fig:sym_schematic}). Because
$B_z$ should not change its sign over the equatorial plane,
the symmetry in $b_{R, \varphi}$ leads to the anti-symmetry
in field line morphologies and vice versa.  For symmetric
modes, we set the mid-plane values
$\delta b_{R 0} = \delta b_{\varphi 0} = 0$,
$\delta j_{R 0} = 1$, but $\delta j_{\varphi 0}$ as the
other free parameter (the extra subscripts ``0'' here denote
the $\zeta=0$ values). For anti-symmetric modes, we fix
$\delta j_{R 0} = \delta j_{\varphi 0} = 0$,
$\delta b_{R 0} = 1$, but $\delta b_{\varphi 0}$ as the
other paramter.  When obtaining these type of modes,
parameters $\nu$ and $\delta j_{\varphi 0}$ (symmetric) and
$\delta b_{\varphi 0}$ are adjusted, so that
$\delta b_R |_{\zeta = \zeta_{\rm wb}} = \delta b_{\varphi}
|_{\zeta = \zeta_{\rm wb}} = 0$.  This approach of setting
up boundary conditions constrains the perturbations that
they do not introduce shears in magnetic fields at
$\zeta_{\rm wb}$, to avoid electric current sheets with
infinitesimal thickness causing discontinuities. It also
guarantees the solenoidal conditions of the perturbation
fields and the radial locality of the modes. We refer the
reader to Appendix~\ref{sec:apdx-boundary} for the
elaboration and analyses on these boundary conditions.


\section{Stabilities of Simplified Models}
\label{sec:simp-modes}

This section applies the methods in \S\ref{sec:method} to
simplified systems, which verify that our theories comply
with existing studies, and also help us to acquire
acquaintances with the behavior of the perturbations in
PPDs.

\subsection{Local ideal MHD limit}
\label{sec:simp-ideal-mhd}

The perturbation theories derived in
\S\ref{sec:linear-perturb} quantify the finite
conductivities of plasmas with non-zero dimensionless
diffusivity $\alpha$, and can thus be restricted to the
ideal MHD limit by taking $\alpha\rightarrow 0$, yielding,
\begin{equation}
  \begin{split}
  & \partial_{\tau} \delta \mu_R = \frac{2 \partial_{\zeta}
    \delta b_R}{\beta_0 \varrho} + 2 \delta \mu_{\varphi} , \
    \partial_{\tau} \delta \mu_{\varphi}
  = \frac{2\partial_{\zeta} \delta b_{\varphi}}{\beta_0
    \varrho} - \frac{\delta \mu_R}{2} , \\
  & \partial_{\tau} \delta b_R = \partial_{\zeta} \delta
    \mu_R ,\
    \partial_{\tau} \delta b_{\varphi} = \partial_{\zeta}
    \delta \mu_{\varphi} - \frac{3}{2}  \delta b_R . 
  \end{split}
\end{equation}
At any specific altitude, $\beta_0 \varrho$ is fixed. One
usually studies the local behaviors by assuming that the
coefficients can be treated as constants and Fourier
transforms are applicable, i.e.,
$\delta x\propto \e^{\i(k\zeta - \omega t)}$, and the
dispersion relation can be written as,
\begin{equation}
  \label{eq:ideal-dispersion-mat}
  \begin{bmatrix}
  \i \omega &  & \i k & \\
   -3/2 & \i \omega &  & \i k\\
    2\i k / (\beta_0\varrho) &  & \i \omega & 2\\
  & 2\i k / (\beta_0\varrho) k & -1/2 & \i \omega
  \end{bmatrix}
  \begin{bmatrix}
    \delta b_R\\
    \delta b_{\varphi}\\
    \delta \mu_R\\
    \delta \mu_{\varphi}
  \end{bmatrix} = 0 .
\end{equation}
This equation requires a vanishing determinant for
non-trivial solutions. Defining
$\tilde{k}^2 \equiv 2 k^2 / (\beta_0 \varrho)$, the
dispersion relation reads,
\begin{equation}
  \label{eq:ideal-dispersion}  
  \omega^4 - (2 \tilde{k}^2 + 1) \omega^2 - 3 \tilde{k}^2
  + \tilde{k}^4 = 0 .
\end{equation}
Since $\omega^2$ is always real, instability arises from
positive imaginary parts of $\omega$, which requires
\begin{equation}
  \label{eq:ideal-dispersion-unstable}
  \omega^2<0\ \Rightarrow\
  2\tilde{k}^2+1 < (16\tilde{k}^2 + 1)^{1/2}\ ,
\end{equation}
or $\tilde{k} < \sqrt{3}$.  This result is
identical to the MRI dispersion relation assuming
isothermal, radial local (ignoring all radial derivatives
and wavenumbers) conditions \citep [e.g.][]
{1991ApJ...376..214B}.

\subsection{Constant diffusivities}
\label{sec:simp-const-diff}
 
\begin{figure*}
  \centering
  \includegraphics[width=7.1in, keepaspectratio]
  {\figdir/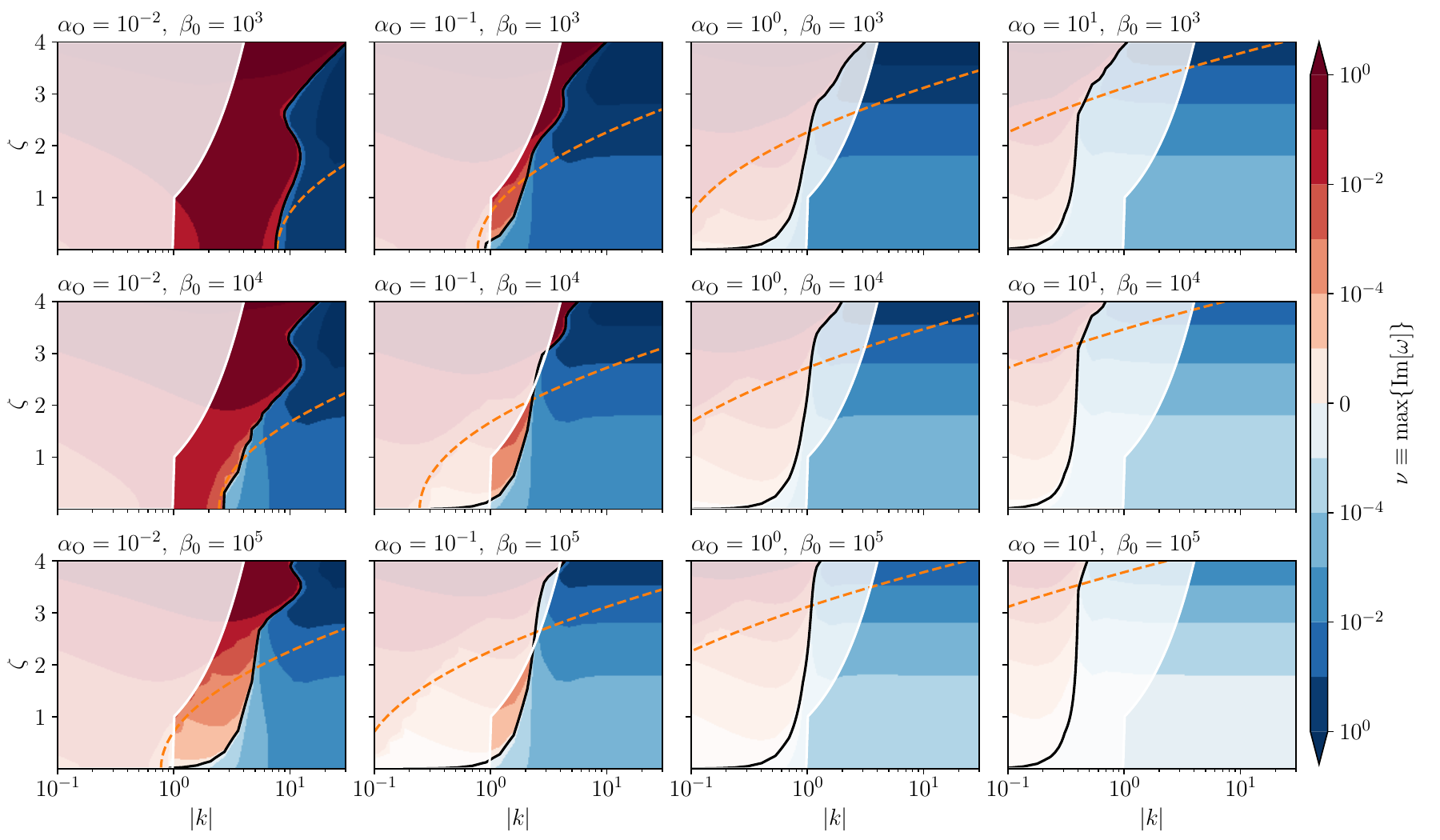}
  \caption{Local-mode growth rates
    ($\nu \equiv \max\{{\rm Im} [\omega]\}$) assuming
    constant diffusivities, showing the distribution in the
    space spanned by the modulus of dimensionless wavenumber
    $|k|$ and the dimensionless altitude $\zeta$
    (\S\ref{sec:simp-const-diff}). The colormaps are in
    symmetric logarithmic scales: red for positive $\nu$,
    and blue for negative $\nu$, separated by solid black
    curves indicating the $\nu = 0$ contours. Dashed curves
    in orange indicate the limiting critical wavenumber
    $k = (3\xi)^{1/2}/\alpha_\O $ derived in
    \citet{1999ApJ...515..776S} for the weakly ionized disk
    with magnetic Reynolds number $R_{\rm m}\ll 1$.  The
    dimensionless diffusivity $\alpha_\O$ and the plasma
    $\beta$ of models are presented to the upper-left of
    each panel. \response{The shaded regions on the
      upper-left side of the white solid curves mark the
      regions where $k \leq \partial_\zeta\ln\varrho$ or
      $k \leq 1$. Local calculations of the dispersion
      relation are reasonable only when
      $k\gg \partial_\zeta \ln\varrho$ and $k > 1$, which
      does {\it not} hold within this shaded region.}  }
  \label{fig:dispersion_const_diff}
\end{figure*}

A more meaningful simplification is to conduct calculations
with constant, non-zero diffusivities. Similar to the ideal
MHD case, we insert the perturbations into
eq.~\eqref{eq:method-perturb-ode} using the vertically local
approximation and the Fourier transform scheme. Assuming
that all
dimensionless diffusivities ($\alpha_{\O,\H,\A}$) are
constants of $\zeta$, the equations reduce to
\begin{equation}
  \label{eq:simp-perturb-ode-const}
  \begin{split}
  & \begin{bmatrix}
   \i\omega &  & a_{\varphi R} & a_{\varphi\varphi} \\  
     -3/2 & \i\omega & a_{RR} & a_{R\varphi} \\
    & \i k & 1 & \\
    -\i k &  &  & 1
    \end{bmatrix} 
   \begin{bmatrix}
    \delta b_R\\
    \delta b_{\varphi}\\
    \delta j_R\\
    \delta j_{\varphi}
   \end{bmatrix} = 0\ ;
    \\
    & a_{\varphi R} \equiv  \i k\alpha_{\varphi R} - 2\xi(
    \zeta + \i k ) ,\\
    & a_{\varphi\varphi} \equiv \i k\alpha_{\varphi\varphi}
      + \xi\omega(k - \i\zeta)  ,\\
    & a_{RR} \equiv -\i k\alpha_{RR} - \xi\omega(k 
      - \i\zeta)  ,\\
    & a_{R\varphi} \equiv -\i k\alpha_{R\varphi} -
      \xi(\zeta + \i k)/2 .
  \end{split}
\end{equation}
Zero determinant of the matrix yields the dispersion
relations. It is straightforward to verify that the
dispersion relation reduces to
eq.~\eqref{eq:ideal-dispersion} once
$\alpha_{\O,\H,\A}\rightarrow 0$ and $\zeta\rightarrow
0$. An extra fact allows us to simplify the analyses
further, that one shall have $\alpha_\O\gg \alpha_\H$ and
$\alpha_\O\gg \alpha_\A$ inside most regions below the disk
surfaces \citep[e.g.][]{2016ApJ...819...68X}, leading to,
\begin{equation}
  \label{eq:simp-perturb-alpha-simp}
  \alpha_{RR}\sim\alpha_{\varphi\varphi}\sim\alpha_\O ,
  \ \{ \alpha_{RR}, \alpha_{\varphi\varphi}\} \gg
  \{ \alpha_{\varphi R}, \alpha_{R\varphi}\}  .
\end{equation}

We use $\alpha_\O$ and $\beta_0$ as the parameters of
models. Similar to \S\ref{sec:simp-ideal-mhd},
Figure~\ref{fig:dispersion_const_diff} plots
$\nu \equiv \max\{{\rm Im}[\omega]\}$ in the space spanned
by the modulus of wavenumber $|k|$ and $\zeta$ for models
with different $\alpha_\O$ and $\beta_0$. \response{It is
  noted that $k \lesssim 10^0$ is unphysical in disks since
  the surfaces of a typical PPD is usually 2 to 3 scale
  heights from the equatorial plane. In addition, local
  calculations of the dispersion relation are only valid on
  spatial scales smaller than the variation lengths for
  typical disk parameters, viz.
  $k \gg \partial_\zeta \ln \varrho$. For a complete
  presentation, we still analytically prolongate our
  solutions to $|k| = 10^{-1}$, and mark the invalid regions
  with white shades.}

In Figure~\ref{fig:dispersion_const_diff} we observe that,
at each altitude there is a critical $k_\crit$ that splits
$\nu > 0$ and $\nu < 0$. The location of this $k_\crit$ is
mostly insensitive to $\beta_0$, and even insensitive to
$\zeta$, but is related to $\alpha_\O$.  At the
$\zeta \rightarrow 0$ limit, the dispersion relation reduces
to
\begin{equation}
  [\omega(1+k^2\xi) + \i \alpha_\O k^2 ]^2 + 3k^2\xi = 0
\end{equation}
If one further assumes that $k^2 \xi \ll 1$ (that mostly
holds when $\beta_0\gg 1$ at $\zeta = 0$) and
$|\omega|^2 \ll 1$ (so that $\xi = 2/\beta_0\varrho$), which
follows the magnetic Reynolds number
$R_{\rm m} \equiv \xi(1-\omega^2) \alpha_\O^{-1} \ll 1$,
such dispersion relation is identical to the weakly ionized
Keplerian disk perturbations described in
\citet{1999ApJ...515..776S}. Such limiting case yields the
highest unstable wavenumber
$k_{\rm crit} = (3\xi)^{1/2}/\alpha_\O $ and the most
unstable wavenumber $k_{\rm m.u.} = k_\crit / 2$.  However,
considering the cases where $k^2\xi \ll 1$ or
$\xi(1-\omega^2) \alpha_\O^{-1} \ll 1$ {\it no longer}
holds, the actual $k_\crit$ could differ significantly from
the previous scenarios, as one can observe by comparing the
black contour (for actual $k_\crit$) and the white dashed
curve (for $(3\xi)^{1/2}/\alpha_\O$) in
Figure~\ref{fig:dispersion_const_diff}. Numerically we fit
and identify that the critical $k$ roughly follows a simple
power-law, $k_{\crit}\sim \alpha_\O^{-1/2}$ at relatively
low altitudes ($10^{-1} \lesssim \zeta \lesssim 2$).

When $k_{\crit}$ is comparable to unity, the unstable modes
can grow, as the modes with vertical wavenumbers smaller
than $2\pi/h$ do not exist. This criterion about the growth
of instabilities agrees with the conclusions in e.g.,
\citet{2007ApJ...659..729T} and \citet{2008A&A...483..815I}
semi-quantitatively, although it is noteworthy that this
result relies on the constant-diffusivity assumption.
 
\begin{figure*}
  \centering
  \includegraphics[width=7.1in, keepaspectratio]
  {\figdir/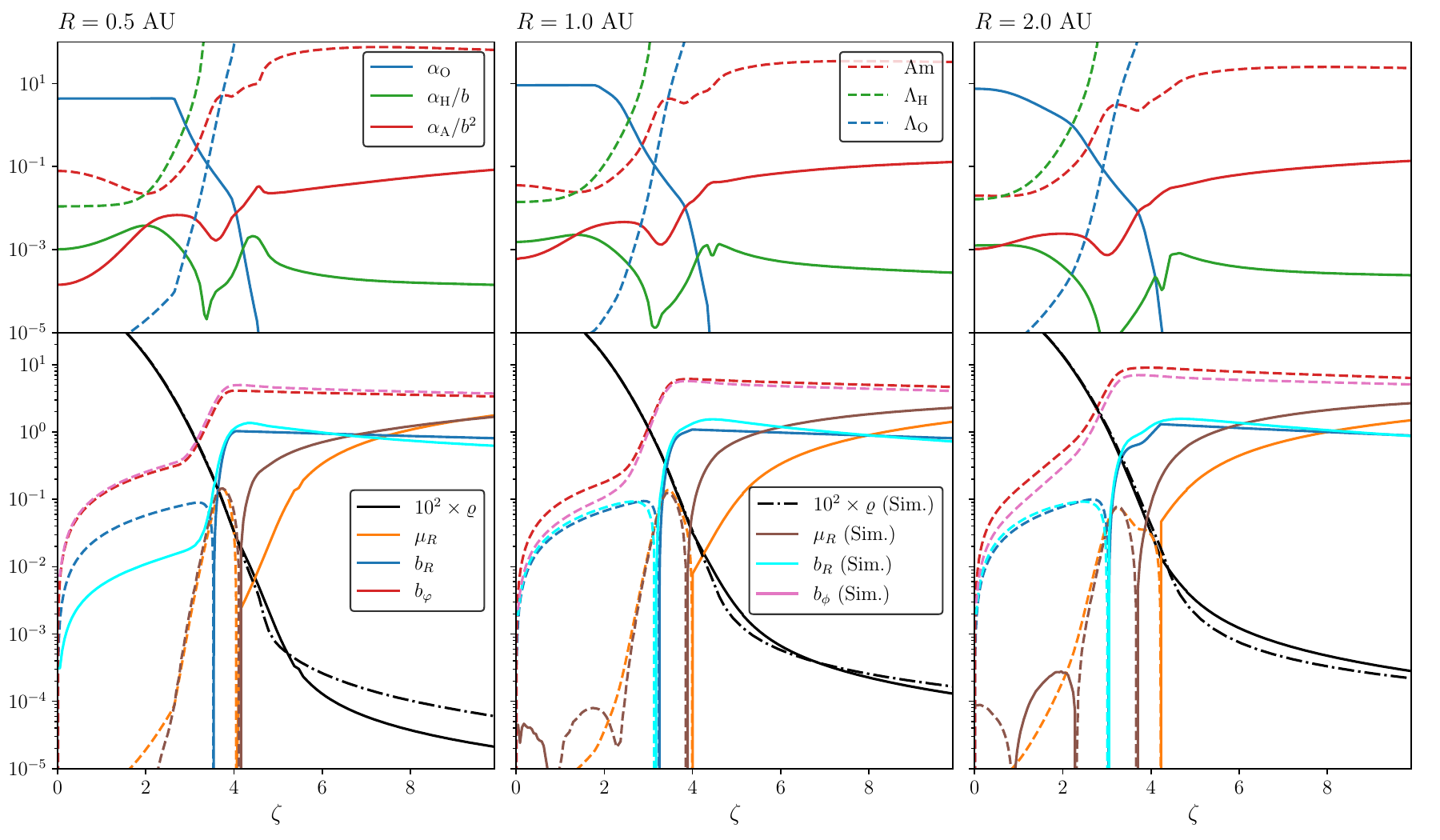}
  \caption{Non-ideal MHD profiles for thei fiducial model
    (\S\ref{sec:fid-steady}) at $(R/\au)\in\{0.5, 1, 2\}$
    presented in different columns. {\bf Top row}: the
    dimensionless diffusivities ($\alpha_\O$, $\alpha_\H/b$,
    and $\alpha_\A/b^2$) and Elsasser numbers ($\Lambda_\O$,
    $\Lambda_\H$, ${\rm Am}$). {\bf Bottom row} compares the
    key MHD profiles ($\varrho$, $b_R$, $b_\varphi$,
    $\mu_R$) of the semi-analytic solutions to the
    simulation results [marked with ``(Sim.)''] used in
    \citet{2020ApJ...904L..27N},
    \citet{2023NatAs...7..905F}. Note that in the bottow
    row, different physical quantities are distinguished by
    colors, while dashed and solid line shapes indicate
    negative and positive values, respectively. }
  \label{fig:fiducial_mhd_profiles}
\end{figure*} 

\section{Fiducial Model: Magnetic Diffusivites Calculated
  by Simulations}
\label{sec:fid-model}

In realistic magnetized protoplanetary disks (PPDs) models,
the diffusivity profiles are vertically stratified, with a
dynamical range of several orders of magnitude. In contrast
to more straightforward local modes discussed in section
\S\ref{sec:simp-modes}, these profiles require solving
eigenvalue problems (as per eq.~\ref{eq:method-perturb-ode})
using semi-analytic methods described in section
\S\ref{sec:linear-perturb}. This section's calculations rely
on the non-ideal MHD profiles observed in the fiducial
simulation adopted by \citet{2020ApJ...904L..27N} and
\citet{2023NatAs...7..905F}.

\subsection{Comparisons of steady-state solutions}
\label{sec:fid-steady}

In order to verify the validity of our semi-analytic
approach, we compare the simulation profiles with the
steady-state solutions yielded by the scheme described in
\S\ref{sec:method-steady}, taking the vertical distributions
of magnetic diffusivities at different radii as the input.
It is noticed that the $|b_R|\ll 1$ near the mid-plane,
where small variations may not affect the overall behavior
of the solution, yet will still change the apparent profiles
in the comparisons. Here we notice that, if we define
$\chi_z\equiv \partial_{\ln R}\ln B_R$, then,
\begin{equation}
  \partial_{\zeta} b_R = j_{\varphi} + \frac{\chi_z}{\mu_k} ,
\end{equation}
and the $\chi_z/\mu_k$ term will be rather important near
the equatorial plane where
$|j_\varphi|\simeq |\partial_\zeta b_R| \ll 1$ and
$|b_R|\ll 1$. A reasonable choice is $\chi_z = -1$, which
leads to $R^{-1}\partial_R(R B_R) = 0$, and
$\partial_z B_z = 0$ is naturally equivalent to
$\nabla \cdot \B = 0$. This choice also fits the
$R\gtrsim 1~\au$ regions in the simulations reasonably well,
as it corresponds to a $B_{z0} \propto R^{-1}$ profile which
largely approximates the initial magnetic fluxes at the
equatorial plane. Therefore, unless specially indicated, we
use $\chi_z = -1$ in what follows.


Figure~\ref{fig:fiducial_mhd_profiles} presents the
comparisons to the simulation results at
$(R/\au)\in \{0.5, 1, 2\}$, confirming that our
semi-analytic approach indeed yields consistent results. The
steady-state solutions should be sufficiently regular at
different radii when used as the foundation of perturbation
theories. At the same time, our numerical tests have found
that varying $\chi_z$ will {\it not} have visible impacts on
the eigenvalues.
 
\begin{figure*} 
  \centering
  \includegraphics[width=7.1in, keepaspectratio]
  {\figdir/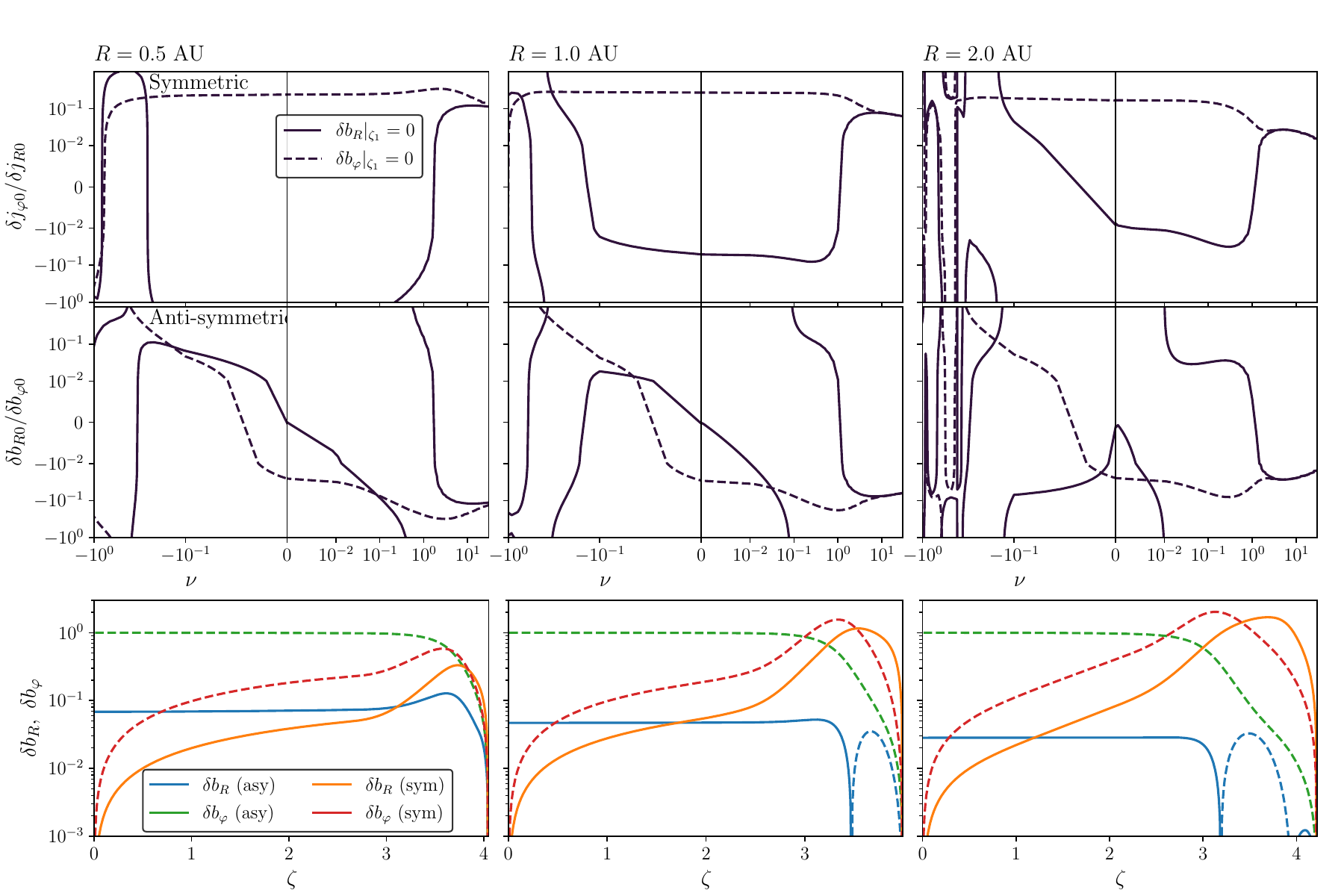}
  \caption{Eigensolutions at disk radii
    $(R/\au)\in\{0.5, 1, 2\}$ presented in different
    columnes (see also \S\ref{sec:fid-eigen-fixed}). The
    {\bf top row} shows the phase diagrams for the symmetric
    perturbation modes in the space spanned by the
    dimensionless
    $\{\nu_{\rm sym}\}\times\{\delta j_{\varphi 0}/\delta
    j_{R 0}\}$. The solid and dashed contours indicate the
    tracks of $\delta b_R|_{\zeta_1} = 0$ and
    $\delta b_\varphi|_{\zeta_1} = 0$ ($\zeta_1$ is the
    altitude of the wind base), respectively, and their
    intersections indicate eigensolutions. Note that the
    negative-$\nu$ and positive-$\nu$ regions, separated by
    a vertical solid line, have different scales in
    $\nu$. The {\bf middle row} is similar to the top row
    but shows the phase diagrams for the anti-symmetric
    modes in the
    $\{\nu_{\rm asy}\}\times\{\delta b_{R0}/\delta
    b_{\varphi }0 \}$ phase space. The {\bf bottom row}
    presents the eigensolutions with the largest growth
    rates for both modes. Note that the symmetric modes
    always have negative growth rates, and that the tracks
    seem to converge at large positive $\nu$ but they never
    cross each other indeed. }
  \label{fig:fiducial_phase} 
\end{figure*} 

\subsection{Eigenmodes of perturbations}
\label{sec:fid-eigen}

Once the steady-state solutions have been obtained, the
growth rates of perturbation modes can be calculated. For
the symmetric modes, adjusting two parameters is necessary
to make $\delta_{R,\varphi}$ vanish simultaneously at $\zeta
= \zeta_1$, the top of the accretion layer. Because the
perturbative problems are linear, one can always set $\delta
j_{R0} = -[\partial_\zeta \delta b_\varphi]_{\zeta = 0} =
1$, and treat $\delta j_{\varphi 0} = [\partial_\zeta \delta
b_R]_{\zeta = 0}$ as one of the parameters. For the
anti-symmetric modes, one sets $\delta b_{R0} = 1$, and
$\delta b_{\varphi 0}$ is the parameter. The other parameter
in both cases is always the eigenvalue $\nu$.

\subsubsection{Eigenmodes at various disk radii }
\label{sec:fid-eigen-fixed}

We elaborate the calculations for $R/\au\in \{0.5, 1, 2\}$
as examples. Plotting the track of the zero points of
$\delta b_{R1} \equiv \delta b_R|_{\zeta_1}$ and
$\delta b_{\varphi 1}\equiv \delta b_\varphi|_{\zeta_1}$ in
the phase space spanned by
$\{\nu\} \otimes \{\delta j_{\varphi 0}/\delta j_{R 0}\}$
(symmetric modes) or
$\{\nu\} \otimes \{\delta b_{R 0}/\delta b_{\varphi 0}\}$
(anti-symmetric modes), the proper eigen modes should locate at
the intersections of the ``root'' curves for $\delta b_{R1}$
and $\delta b_{\varphi 1}$ 
(Figure~\ref{fig:fiducial_phase}).

The first attempts to obtain eigenmodes intentionally set
$\alpha_\H = 0$ as the fiducial cases. For both types of
modes, we observe that the modes with $|\nu| \lesssim
10^{-1}$ have the least number of nodes; modes with higher
spatial frequencies all lie in the $\nu < 0$ half-space. In
the $\nu > 1$ domain, the traces of roots seem to converge,
but more detailed examinations confirm that they never touch
each other.

At this radius, the maximum $\nu$ for symmetric modes is
$\nu \simeq -0.35 < 0$ for $R = 1~\au$. Recovery of
dimensions finds that this mode decays at $\sim 0.4~\yr$ per
e-fold. This value is tiny compared to the disk lifetime
($\gtrsim 10^6~\yr$), let alone other symmetric modes with
higher spatial frequency and more negative $\nu$. Therefore,
no instability will occur through the symmetric modes. In
contrast, an anti-symmetric mode with
$\nu \simeq 0.0337 > 0$ exists. Amplitudes of this growing
mode will increase by $\sim 5~\yr$ per e-fold and should
leave the linear stage rather quickly.

\response{With the current choice of magnetic diffusivity
  parameters, it is noteworthy that the Hall effect does not
  efficiently manipulate this instability. Using the
  $R = 1~\au$ vertical Hall diffusivity profile
  ($\alpha_\H/b$) obtained based on the thermochemical
  profiles in \citet{2020ApJ...904L..27N}, we test the
  situations that $B_z$ is parallel and anti-parallel to the
  axis of disk rotation, respectively.} The resulting
eigenvalues are affected by a tiny fraction ($\nu = 0.0343$
for parallel fields and $\nu = 0.0332$ for anti-parallel
fields). This consequence seems to be in odd to some
previous studies concluding that the Hall effect could lead
to significant asymmetries across the equatorial plane
\citep[e.g.][] {2014A&A...566A..56L,
  2017ApJ...836...46B}. \response{ We point out that the
  apparent discrepancy originates from the distinct magnetic
  diffusivity profiles selected for the disk model. The
  profiles utilized in the work of
  \citet{2020ApJ...904L..27N} result in a ratio of
  $|\alpha_\H/b|/\alpha_\O$ that is less than $10^{-1}$ (and
  frequently even $\lesssim 10^{-3}$) beneath the disk
  surfaces (see also
  Figure~\ref{fig:fiducial_mhd_profiles}). It is essential
  to recognize that the actual magnetic diffusivity profiles
  are contingent upon the thermochemical conditions at hand,
  which result from the specific thermochemical network
  employed and the astrophysical conditions of the concerned
  regions, such as the high-energy radiation luminosity
  emitted by the central star. The Hall diffusivity profiles
  that have been integrated into studies concentrating on
  the Hall shear instability and subsequent symmetry
  breaking \citep[e.g.][] {2014A&A...566A..56L,
    2017ApJ...836...46B, 2024MNRAS.530.5131S} are typically
  significantly higher. } Once
$|\alpha_\H/b| \gtrsim \alpha_\O$, local analyses similar to
\S\ref{sec:simp-const-diff} reveals that the modes become
significantly more unstable: the critical wavenumber of
instability becomes $\gtrsim 10\times$ greater than the
current values.  \response{In the meantime, even the results
  assuming $\alpha_\H/b\ll \alpha_\O$ do {\it not} deny the
  importance of the Hall effect.} The nature of the
spontaneous symmetry breaking can the Hall effect could be
the ``first push'' that determines the direction, and the
subsequent amplification of asymmetry is dominated by the
instabilities described in this section.

\subsubsection{Scalings of eigenvalues}
\label{sec:fid-eigen-scale}

\begin{figure}
  \centering
  \hspace*{-0.4cm} 
  \includegraphics[width=3.6in, keepaspectratio]
  {\figdir/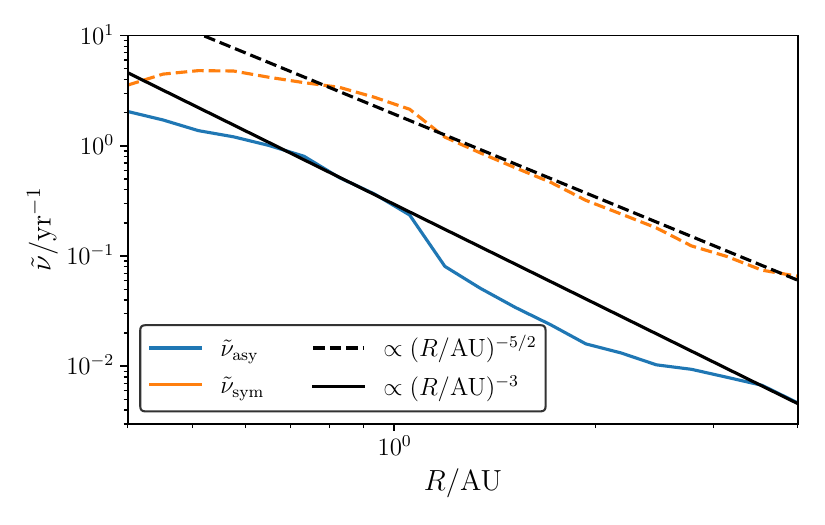}
  \caption{Dimensional growth rates of anti-symmetric
    ($\tilde{\nu}_{\rm asy}$) and symmetric
    ($\tilde{\nu}_{\rm sym}$) modes (dashed line for
    negative values), for the fiducial model through the
    range of disk radius $0.4<(R/\au)<4$.  Note that
    $\tilde{\nu}_{\rm sym} < 0$ holds for all radii. The
    lines indicating the $R^{-5/2}$ and $R^{-3}$ power-laws
    are also presented for comparison.}
  \label{fig:fiducial_var_R}
\end{figure}

When the scheme described in \S\ref{sec:fid-eigen-fixed} is
applied to all relevant radii in the disk model in
\S\ref{sec:fid-steady}, stability conditions of the disk can
be studied in terms of the reflection symmetry.  After
multiplied by the local Keplerian angular velocity to
recover the dimensions, we present the dimensional growth
rates as functions of disk radius in
Figure~\ref{fig:fiducial_var_R}--$\tilde{\nu}_{\rm sym}$ for
the symmetric mode, and $\tilde{\nu}_{\rm asy}$ for the
anti-symmetric mode. It is evident that throughout all radii
in the model $\tilde{\nu}_{\rm sym} < 0$ but
$\tilde{\nu}_{\rm asy} > 0$, thus symmetric modes always
decay, but at least one of the anti-symmetric modes
grows. In other words, the reflective symmetric morphology
of the wind-driven accretion model is unstable, and the
instability always initiates in an anti-symmetric shape.

For both $\tilde{\nu}_{\rm asy}$ and the
$|\tilde{\nu}_{\rm sym}|$ at $R \gtrsim 0.6~\au$, one can
roughly read their scaling from
Figure~\ref{fig:fiducial_var_R}, 
\begin{equation}
  \label{eq:fid-nu-scaling}
  \dfrac{|\tilde{\nu}_{\rm sym}|}{\yr^{-1}} \sim 2 \times
  \left(\dfrac{R}{\au} \right)^{-5/2},\ \ 
  \dfrac{\tilde{\nu}_{\rm asy}}{\yr^{-1}} \sim 0.2 \times  
  \left(\dfrac{R}{\au}\right)^{-3} .
\end{equation}
The growth and decay rates decrease with radius faster than
$\Omega_\K$. For larger radii, the growth of anti-symmetric
modes could be slow. If the scaling relation in
eq.~\eqref{eq:fid-nu-scaling} can be generalized to all
concerned radii in a typical PPD, then
$\tilde{\nu}_{\rm asy}^{-1}$ could be as slow as
$\sim 2.5\times 10^6~\yr$ at $R\sim 200~\au$, presumably the
``typical'' outer radius of PPDs: the anti-symmetric
instability will marginally grow through the
$\sim 10^6-10^7~\yr$ lifetime of a PPD. The growth is
challenging to identify in numerical simulations available
at such a large distance, and hence, the necessity of the
analytic approach is emphasized.

\subsection{Evolution of the anti-symmetric modes}
\label{sec:fid-evolution}

For the breaking of reflection symmetry, we have identified
that a possible evolution manner is a quasi-steady state,
whose most prominent characteristic is the asymmetric
wind-driven accretion flow. To construct the morphologies
analytically, we slightly modify the methods in
\S\ref{sec:method-steady} by setting non-vanishing $b_R$ and
$b_\varphi$ at the mid-plane and integrating to both
$+\zeta$ and $-\zeta$ directions. When obtaining such
models, the matching onto ideal MHD wind solutions is
carried out only on the wind-launching side, and the other
side is left free. This scheme will leave the mid-plane
$b_{R0}$ and $b_{\varphi0}$ unconstrained. These two degrees
of freedom reserve the tracks of growth for the
anti-symmetric modes, allowing a selection of $(b_{R0},
b_{\varphi 0})$ reflects a snapshot in the path of growth
before saturation. In general, the evolution of
instabilities could be a complicated issue that requires
excessive numerical experiments to understand, and we leave
these studies to future works.

\begin{figure}
  \centering
  \hspace*{-0.4cm} 
  \includegraphics[width=3.5in, keepaspectratio]
  {\figdir/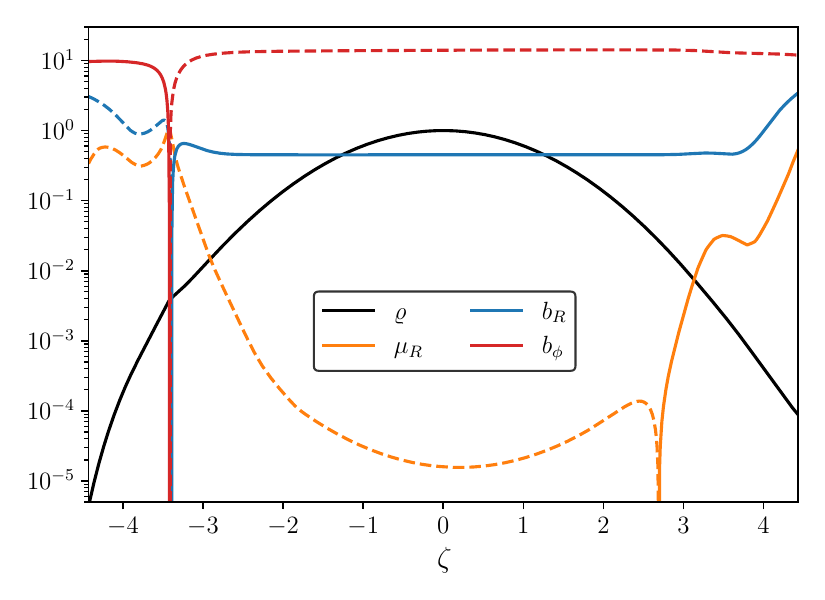}
  \caption{ Similar to the lower row of
    Figure~\ref{fig:fiducial_mhd_profiles} but showing a
    non-symmetric solution using the diffusivity profiles at
    $R=1~\au$ as an example. Colors distinguish different
    physical quantities, while the line shapes indicate the
    signs (solid for positive, dashed for negative).}
  \label{fig:fiducial_double_sided}
\end{figure}

Figure~\ref{fig:fiducial_double_sided} illustrates the MHD
profiles obtained at $R=1~\au$, in which we set
$b_\varphi|_{\zeta=0} = -14$, and the mid-plane value of
$b_R$ is fixed at $b_R|_{\zeta=0} = 0.45$. In this model,
the accretion mainly occurs on the $\zeta < 0$ side, while
the disk wind is launched efficiently only on the
$\zeta > 0$ side. Over the disk surface on the $\zeta < 0$
side, suppression of gas density is observed, and $\mu_R$ is
always negative. Such a combination of fluid parameters
indicates an absence of wind over that surface.

The one-sided wind launching feature implies that the disk
wind will exhibit an apparent ``asymmetric'' shape. Such
phenomenon has been observed in various non-ideal MHD
simulations of PPDs, in which the asymmetries developed to
different extents (e.g. \citealt{2017ApJ...845...75B};
\citealt{2019ApJ...885...36H};
\citealt{2020A&A...639A..95R};
\citealt{2022MNRAS.516.2006H}; X. Hu et al., in
prep.). Calculations in this work have confirmed that these
asymmetric morphologies should be physically plausible, and
we should also expect to discover such asymmetries
observationally (see discussions in
\S\ref{sec:discuss-obv}).
 
\section{Stability in Various MHD Models}
\label{sec:var-models}

The wind-driven accretion problem described in \S ef
{sec:fid-model} is a typical PPD model, yet countless
situations still need to be discussed. The diversity of
PPDs, including their mass distributions, host star
properties, and thermochemical conditions, can all affect
the diffusivity profiles significantly. To study those
parameters, this section explores different disk conditions,
including gravitation, magnetization, and diffusivity
profiles. We will manipulate one parameter at a time to
simplify and clarify the elaborations.

\subsection{Functional forms of the vertical diffusivity profiles}
\label{sec:var-functional}

For typical PPDs, the vertical stratification of
diffusivities can be qualitatively divided into three
regions, (1) the weak coupling mid-plane
($\alpha_\O \sim 10^1$) in the $\zeta\lesssim 2$ region, (2)
the transition layer near the disk surface
($10^1\gtrsim \alpha_\O\gtrsim 10^{-4}$,
$2 \lesssim \zeta/h \lesssim 4$), and (3) the highly ionized
wind above. As we have verified that the Hall effect is more
critical as a ``first-push'' than an amplification
mechanism, the following discussions will focus on the Ohmic
and ambipolar diffusivities. We choose to parameterize
$\alpha_\O$ and $\alpha_\A/b^2$, as they are insensitive to
magnetic field strengths.  Simple log-linear functions with
caps parameterize the vertical profiles of diffusivities
($\zeta' \equiv \max\{(\zeta - \zeta_t),\ 0\}$),
\begin{equation}
  \label{eq:var-functional-alpha}
  \begin{split}
    & \log_{10}\alpha_\O = \log_{10}\alpha_{\O 0} + \psi_\O
      \,\zeta' ,\\
    & \log_{10}(\alpha_\A/b^2) = \log_{10}(\alpha_\A/b^2)_0
      + \psi_\A\,\zeta' ,
  \end{split}
\end{equation}
where $\alpha_{\O 0}$ and $(\alpha_{\A}/b^2)_0$ are the
values at the equitorial plane, $\psi_\O$ and $\psi_\A$ are
the slopes in the logarithmic space, and $\zeta_t$ marks the
transition altitude. In \S\ref{sec:fid-model} we already
found that $\varrho$ profiles are almost unaffected by the
accretion and remain almost identical to the static disk. In
addition, although the radial derivative of $B_z$ may affect
the $b_R$ near the mid-plane in steady-state solutions, we
have nevertheless confirmed that this hardly affects the
growth rates of perturbation modes. Therefore, we take the
approximations that $\rho = \exp(-\zeta^2/2)$ and
$\chi_z = 0$ (\S\ref{sec:fid-steady}) for all models in this
section to make the results independent of $R$. 

A reference model is constructed and illustrated in the left
column of Figure~\ref{fig:var_model}. This model has
$\alpha_{\O 0} = 10^1$ , $(\alpha_\A/b^2)_0 = 10^{-3}$,
$\psi_\O = -2$, $\psi_\A = 0.5$, and $\zeta_t = 2.5$, which
qualitatively resembles the $R=1~\au$ slice of the fiducial
model in \S\ref{sec:fid-steady}. The consequent steady-state
solution is also quite similar. For the perturbation modes,
the symmetric mode has decay rate
$\nu_{\rm sym}\simeq -0.42$, which has the same sign, but
the absolute value is different by $\sim 20\%$. The
growth rate of the anti-symmetric mode is
$\nu_{\rm asy} = 0.061$, $\sim 50\%$ different from the
results based on the simulation diffusivity profiles.

\subsection{Impacts of MHD parameters}
\label{sec:var-result}

\begin{figure*} 
  \centering
  \includegraphics[height=4.24in, keepaspectratio]
  {\figdir/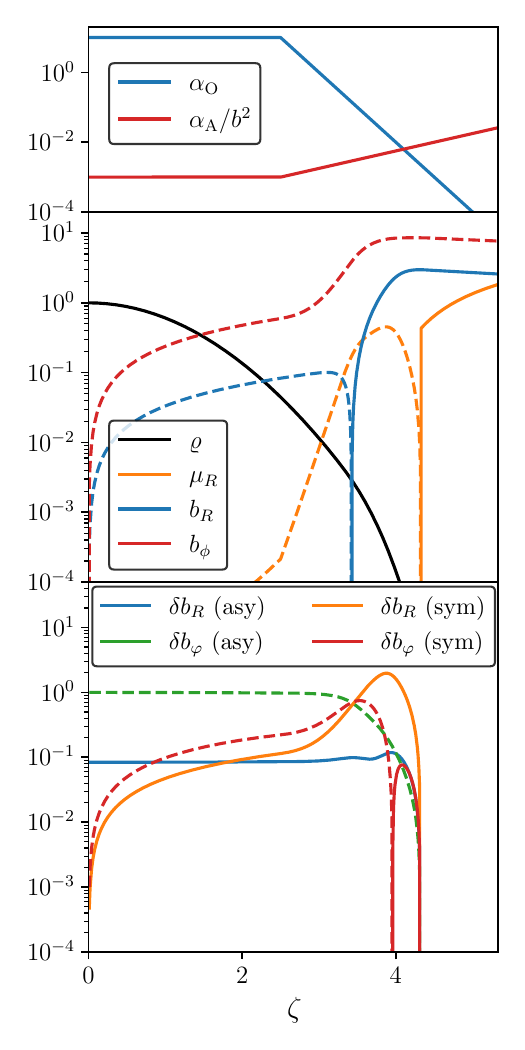}
  \includegraphics[height=4.26in, keepaspectratio]
  {\figdir/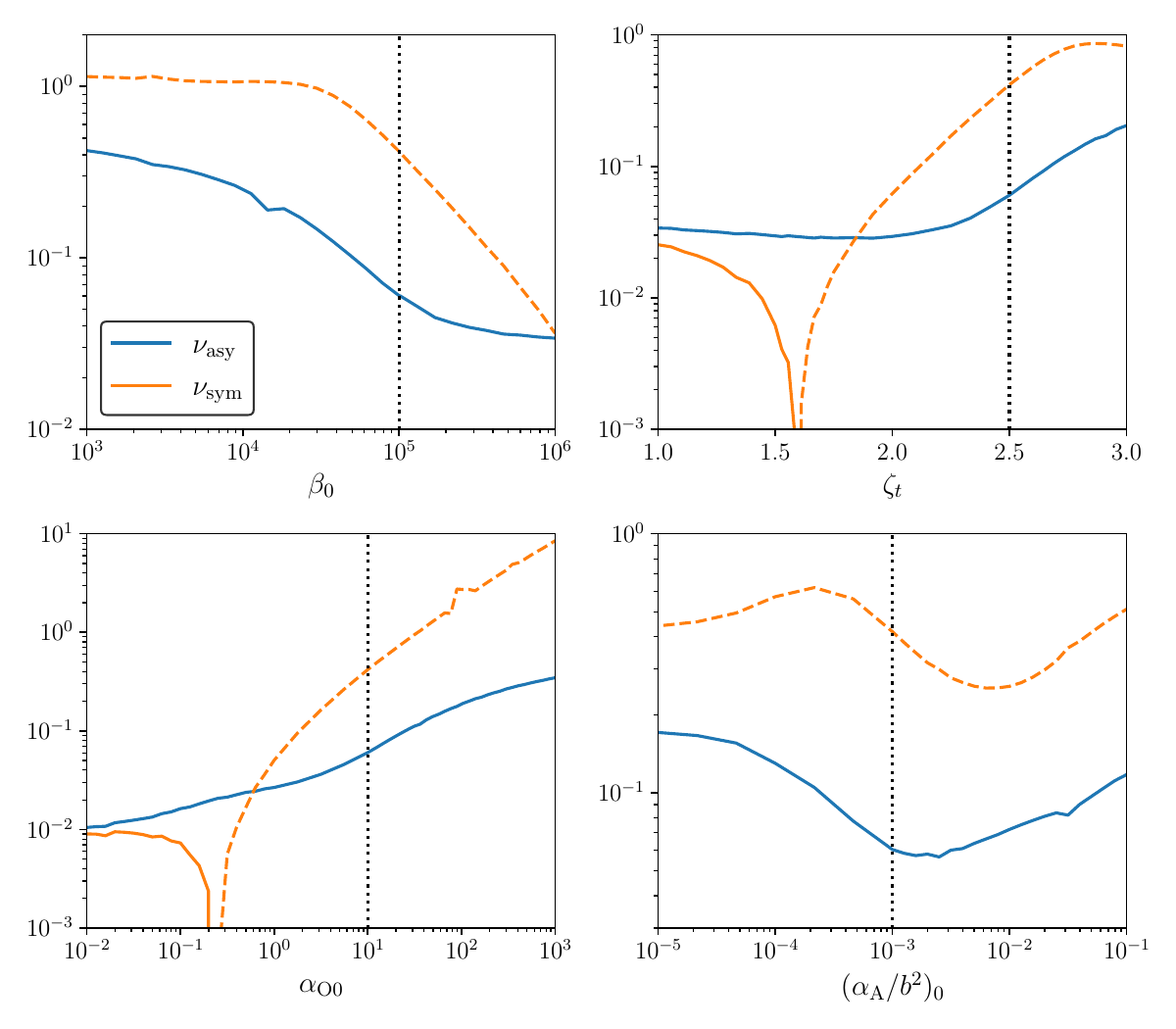}  
  \caption{{\bf Left column} illustrate the model for
    reference with the setup in \S\ref{sec:var-functional}, 
    showing the diffusivity profiles (upper panel), the
    steady-state solution (middle panel), and the
    perturbation modes (lower pannel). The growth rates are
    $\nu_{\rm sym} = -0.10$ and $\nu_{\rm asy} =
    0.035$. {\bf Right two columns} present the dependence
    of $\nu_{\rm sym}$ (orange lines) and $\nu_{\rm asy}$
    (blue lines) on different paramters: $\beta_0$,
    $\zeta_t$, $\alpha_{\O 0}$, and $(\alpha_\A/b^2)_0$ (see
    \S\ref{sec:var-result}). Note that negative values are
    shown in dashed lines. The vertical dotted lines are
    plotted in each panel indicating the value taken by the
    reference model. }
  \label{fig:var_model}
\end{figure*}

The reference model described in \S\ref{sec:var-functional}
is modified to study the influence of different
parameters. We first confirm that $\mu_k$, which reflects
the relative importance of orbital motion, only plays a
secondary role: taking $\mu_k\in [4,\ 10^2]$ only leads to
minor variations in $\nu$, such that $-0.063 \gtrsim
\nu_{\rm sym}\gtrsim -0.10$ and $0.03 \lesssim \nu_{\rm
  asy} \lesssim 0.04$. In contrast, the magnetization
parameterized by $\beta_0$ is a significant parameter. As
shown in Figure~\ref{fig:var_model}, the stronger the disk's
magnetization (or smaller $\beta_0$), the faster the
perturbation modes grow or decay. If one sets $\beta_0 =
10^3$ based on the reference model, such a strong
magnetization yields $\nu_{\rm sym} \simeq -1.0$ and
$\nu_{\rm asy}\simeq 0.38$, leading to an e-fold increase of
anti-symmetric amplitude every half orbit period. With
$\beta_0\lesssim 10^5$, the growth or decay rates scale
roughly as $\sim \beta_0^{-1/2}$, or roughly linearly with
field strength.

Different components of magnetic diffusivities control the
stability problem in different ways. The ambipolar diffusion
parameter $(\alpha_\A/b^2)_0$ affects both $\nu_{\rm sym}$
and $\nu_{\rm asy}$, yet the absolute values of both $\nu$
vary rather reluctantly as the parameter stay in the range
$(\alpha_\A/b^2)_0\in [10^{-4}, 10^{-2}]$. The situation is
similar about the dependence of $\nu_{\rm asy}$ upon
$\alpha_{\O 0}$ and $\zeta_t$. When we examine the symmetric
modes, however, it is found that $\nu_{\rm sym}$ becomes
positive when $\alpha_{\O 0}\lesssim 0.6$, or
$\zeta_t\lesssim 1.4$. As the conductivity in the adjacency
of the mid-plane becomes higher, by either lowering the
$\alpha_{\O 0}$ or setting a lower transition altitude
$\zeta_t$, the stronger coupling between field and fluids
eventually leads to the instability even for the symmetric
modes. The threshold value for $\alpha_{\O 0}$ also matches
the vertically local mode calculations in
\S\ref{sec:simp-const-diff} semi-quantitatively. Meanwhile,
$\nu_{\rm asy}$ does not change sign throughout the subset
of parameter space we explored and illustrated in
Figure~\ref{fig:var_model}, neither in the search over a
broader span of the parameter space that is not elaborated
here. This fact indicates the ubiquity of the instability
for the anti-symmetric modes.

\section{Discussions and Summary}
\label{sec:summary}

This study explores the instability that leads to reflection
symmetry breaking over the equatorial planes in
protoplanetary disks (PPDs) that go through wind-driven
accretion. We develop linear perturbation theories using
steady-state solutions and evaluate eigenmodes and their
corresponding eigenvalues with different
symmetries. Symmetric perturbations always decay under
typical MHD profiles, whereas at least one of the
anti-symmetric modes has a positive growth rate. The growth
of anti-symmetric modes leads to tilted disk kinematics,
where the accretion flow occurs on one side, and the disk
wind is launched on the other. We also investigate the
effects of varying magnetic diffusivity profiles on the
reflection symmetry breaking, finding that the anti-symmetry
is most prominent in regions of the disk with poorly ionized
mid-plane where fields and fluids are weakly coupled and
intermediate surface layers where the ionization and
field-fluid coupling are intermediate.

\subsection{Qualitative behavior of eigenvalues} 
\label{sec:var-qual-eigen}

Qualitative discussions on the behaviors of eigenvalues are
necessary to develop physical insights into the findings of
this paper and guide future explorations.  Eq.~\eqref
{eq:method-perturb-matrix} can be recast in the integrated
form,
\begin{equation}
  \label{eq:qual-eigen-mat-intg}
  \left[( A + \xi V_1 ) \delta b\right]_{\zeta_0}^{\zeta_1}
  = V_0 \int_{\zeta_0}^{\zeta_1}\delta b\, \d\zeta .
\end{equation}
Since we are most interested in the modes with the least
nodes, especially the prospectively anti-symmetric modes
with anti-symmetry, the approximation $\nu\ll 1$ can be
safely adopted, hence $\xi \simeq 2/(\beta_0\varrho)$. The
bottom two rows in the matrix form of eq.~\eqref
{eq:qual-eigen-mat-intg} are indeed equivalent to
$\delta j_R = -\partial_\zeta \delta b_\varphi$ and
$\delta j_\varphi = \partial_\zeta \delta b_R$, while the
top two rows should be analysed for $\nu$,
\begin{equation}
  \label{eq:qual-eigen-diff}
  \begin{split}
    &\nu \mathcal{B}_R  = \left[
      \alpha_{\varphi\varphi}\delta j_\varphi   +
      (\alpha_{\varphi R} - 2\xi) \delta j_R 
      \right]_{\zeta_0}^{\zeta_1},\\
    & \nu \mathcal{B}_\varphi = \left[ \alpha_{RR}\delta
      j_R + (\alpha_{R\varphi} - \xi/2) \delta j_\varphi
      \right]_{\zeta_0}^{\zeta_1} + \dfrac{3}{2}
      \int_{\zeta_0}^{\zeta_1}\delta b_{R}\ \d\zeta\ ;\\
    & \mathcal{B}_R \equiv \int_{\zeta_0}^{\zeta_1}\delta
      b_{R}\ \d\zeta - (\xi \delta
      j_\varphi)_{\zeta_0}^{\zeta_1} ,\\ 
    & \mathcal{B}_\varphi \equiv -\int_{\zeta_0}^{\zeta_1}
      \delta b_{R}\ \d\zeta - (\xi\delta
      j_R)_{\zeta_0}^{\zeta_1} . 
  \end{split}
\end{equation}

\subsubsection{Anti-symmetric modes}
\label{sec:var-qual-anti-sym}

For the anti-symmetric modes, the equation for
$\delta\tilde{b}_R$ is typically easier to analyze by
locating $\zeta_0 = 0$ (at which $\delta j_R = \delta
j_\varphi = 0$) and $\zeta_1$ at the first zero point of
$\delta b_R$ (likely the wind base). Without loss of
generality, the linearity of the equations allows one to set
$\delta b_R > 0$ throughout all altitudes in the concerned
modes, which in turn yields $\delta b_\varphi < 0$, $\delta
j_R|_{\zeta_1} = -[\partial_\zeta \delta
b_\varphi]_{\zeta_1} < 0$, and $\delta j_\varphi|_{\zeta_1}
< 0$ due to the geometries of disks and fields. The
coefficient $\mathcal{B}_R$ on is therefore positive
definite, and the sign of $\nu$ depends on the right-hand
side.

The cases that we concern usually have
$|\delta b_\varphi|_{\zeta_1} \gg |\delta b_R|_{\zeta_1}$
due to the amplification of toroidal fields by rotation,
expecting that
$|\delta j_R|_{\zeta_1} \gg |\delta
j_\varphi|_{\zeta_1}$. In the meantime, the wind base
usually have $\alpha_\A/b^2\sim \alpha_\O$, therefore
$|b_\varphi|_{\zeta_1} \gg |b_R|_{\zeta_1}$ usually leads to
$|\alpha_{\varphi R}|_{\zeta_1} \gg
|\alpha_{\varphi\varphi}|_{\zeta_1}$, which have been
confirmed by all cases involved in
\S\ref{sec:fid-model}. These facts focus on the competition
between $\alpha_{\varphi R}$ and $2\xi$ at $\zeta_1$. Since
$b_R$ and $b_\varphi$ have different signs at the wind base
due to the disk geometries and MHD configurations,
$\alpha_{\varphi R}|_{\zeta_1} > 0$ always holds. Our
numerical experiments have found
$\alpha_{\varphi R}|_{\zeta_1} < 2\xi|_{\zeta_1}$ throughout
the fiducial model, which is consistent with the finding
that $\nu > 0$, i.e. there are always unstable
anti-symmetric modes (Figure~\ref{fig:fiducial_var_R}). In
order to eliminate the instability, the inequality
$\alpha_{\varphi R}|_{\zeta_1} < 2\xi|_{\zeta_1}$ has to be
reverted.

\subsubsection{Symmetric modes}
\label{sec:var-qual-sym}

The symmetric modes are easier to analyze by still setting
$\zeta_0 = 0$ but locate $\zeta_1$ at the first zero point
of $\delta j_R = -\partial_\zeta \delta b_\varphi$. With
such selection, it is easy to verify that
$\delta j_{R,\varphi}|_{\zeta_0} > 0$. Also, taking
$\delta b_R > 0$ at $\zeta > 0$ due to linearity, one can
prove that $\mathcal{B}_R$ is still positive. On the
right-hand side,
$[\alpha_{\varphi\varphi} j_\varphi]^{\zeta_1}_{\zeta_0}$ is
definitely negative. This time, the competition of
$\alpha_{\varphi R}$ versus $2\xi$ usually declares the
victory of the former for sufficiently weakly ionized (thus
$\alpha_\O$ is large), weakly magnetized disks: because now
the values at $\zeta_0$ matter, at which $\varrho = 1$, and
$\xi = 2/\beta_0$ is tiny for large $\beta_0$. When the
fluid-field coupling at the mid-plane becomes stronger, the
inequality $\alpha_{\varphi R} < 2\xi$ may no longer hold,
leading to the instability of the symmetric modes. The
explorations in \S\ref{sec:var-result} have already
witnessed this situation.

\subsubsection{Diffusivities and stabilities}
\label{sec:var-qual-diff}

The discussions elaborated above can relate to the physical
picture of electric currents and diffusivities. The
symmetric modes have non-negligible
$|\delta j_{R,\varphi}| = |\partial_\zeta b_{\varphi, R}|$
near $\zeta = 0$, which requires sufficient conductivity to
develop. However, weakly ionized PPDs do not have sufficient
conductivity there, and the highly resistive plasmas at the
mid-plane damp and inhibit the growth of symmetric
modes. Such damping depends on $\alpha_\O$ more than
$\alpha_\A/b^2$, as the former is generally much greater
than the latter at $\zeta\sim 0$, which is consistent with
the emphases on Ohmic resistivity in the stability analyses
stabilities in e.g. \citet{2007ApJ...659..729T} and
\citet{2008A&A...483..815I}.

Anti-symmetric modes, in contrast, bend their field lines at
much higher altitudes (e.g.,
Figure~\ref{fig:fiducial_phase}), where the diffusivities
are sufficient to support the electric currents
required. Ambipolar diffusion is of greater importance at
this time since $\alpha_\O$ decreases drastically near the
disk surfaces. At the same time, the product $|b_R
b_\varphi|$ has much greater absolute values in the term
$\alpha_{\varphi R}\simeq -(\alpha_\A/b^2) b_R b_\varphi$
(note again that $\alpha_\A/b^2$ is primarily independent of
fields and that $\alpha_\H/b$ is negligible due to the lack
of the $|b_Rb_\varphi|$ factor). Qualitatively, this is also
consistent with the emphasis on ambipolar diffusion in the
analyses of disk stabilities in e.g.,
\citet{2011ApJ...736..144B}.

The semi-quantitative assumptions and approximations on
diffusivities involved for the perturbative analyses have
been examined in the elaborations in \S\ref{sec:fid-model}
and \S\ref{sec:var-models}. At lower disk densities and
higher ionization fractions (e.g., magnetized transitional
PPDs; \citealt{2017ApJ...835...59W}), the stability analyses
could lead to different conclusions, converging to the
situations discussed in \S\ref{sec:simp-modes}.

\subsection{Predictions of observations}
\label{sec:discuss-obv}

The anti-symmetric modes in PPDs could result in the absence
of a magnetized wind on one side of the disk
(\S\ref{sec:fid-evolution}). Some young stellar objects
(YSOs) have been observed to launch asymmetric winds over
both disk surfaces, for example, HH 30
\citep{1996ApJ...473..437B, 1999ApJ...516L..95S}. Some other
observational evidence supports asymmetric winds (see a
review in \citealt{2023ASPC..534..567P}). High spectral
resolution [\ion{O}{I}] $6300~\ang$ emission line has been
used to trace the disk winds (e.g.,
\citealt{2013ApJ...772...60R, 2016ApJ...831..169S,
  2018AA...620A..87M, 2018ApJ...868...28F,
  2019ApJ...870...76B, 2023ApJ...945..112F}). The line
profile of [\ion{O}{I}] typically consists of two types of
components: a high-velocity component (HVC) and a
low-velocity component (LVC). While HVCs are produced in
extended jets \citep{2000A&A...356L..41L,
  2000ApJ...537L..49B, 2002ApJ...580..336W}, LVCs most
likely trace MHD winds \citep{2018ApJ...868...28F,
  2019ApJ...870...76B, 2023ApJ...945..112F,
  2023NatAs...7..905F}. Though the majority (70\%) of LVCs
are blueshifted, 21\% of LVCs are red-shifted by more than
$\gtrsim 1~\km~\s^{-1}$ \citep{2023ApJ...945..112F}. These
redshifted LVCs can be explained if the winds are launched
mainly over the disk surface facing away from the
observer. Future observational studies are also desired to
directly reveal the morphologies of disk winds to confirm
the existence of the asymmetries.

 
The ``tilted'' accretion may be related to the asymmetric
feeding of the central protostar, which is potentially
related to the recycling of the accreted materials and the
formation of asymmetric jets observed on various PPDs
\citep[e.g.][]{2000A&A...356L..41L, 2002ApJ...580..959N,
  2023A&A...670A.126F}.  Nevertheless, the innermost
accretion regions and the formation of jets in PPDs are
complicated, and future studies on the jet morphologies of
the jets are required to reveal and confirm the connections
between asymmetries in jets and accretion layers.
Meanwhile, accretion lifted to the surface layers also
reduces the damping of differential rotation and can induce
magnetorotational instabilities (MRI). Once MRI develops in
the surface layer, the magnetized wind will be suppressed,
and the wind launching area shall exhibit a ``truncation''
subsequently. Recent observations have been conducted with
both spatial and spectral resolutions, which figured out the
range of radii for PPD outflows and examined different wind
launching mechanisms
\citep[e.g.][]{2023NatAs...7..905F}. Similar high-resolution
observations are necessary to confirm these theoretical
predictions and understand the complex physical processes
that shape magnetized wind structures. For instance,
observations that identify the wind launching regions with
spatial resolution could establish a direct link between
tilted accretion, possible surface MRI, and the suppression
of wind launching in PPDs. Both theoretical modeling and
high-resolution observations are vital for unraveling the
interplay between magnetic fields, turbulence, and
gravitational instability.

\subsection{Future works}
\label{sec:discuss-future}

This study focuses on examining the vertical modes with
axisymmetry in relation to the wind-driven accretion in
protoplanetary disks. This approach does not account for the
radial dependence of the problem. A possible extension of
this work would involve studying the radial modes, which
play a crucial role in the radial transport of magnetic
fields. Modulated by the non-ideal MHD features, the radial
flux transport and the accretion flow are intertwined and
can result in the formation and movement of disk
substructures such as rings and gaps. Extending the analytic
research on PPD controlled by non-ideal MHD effects, another
prospective area of investigation involves examining the
breaking of the disk axisymmetry by azimuthal modes. The
close relationship between the azimuthal modes and angular
momentum transport in protoplanetary disks can lead to
complex phenomena like spiral arms, vortices, and disk
warps. Through studying these phenomena, we can better
understand how angular momentum transporting is related to
different perturbation modes and how that impacts the
evolution and dispersal of protoplanetary disks.

Existing theories on PPD substructure gaps often attribute
their formation to the influence of existing
planets. Nevertheless, a natural mechanism that can produce
these structures without the influence of readily formed
planets would pave its way to a broader and more generic
application. Hence, the proposed study of co-evolved radial,
azimuthal, and vertical modes could provide crucial insights
into the natural formation of PPD substructures and
subsequent evolution over time. Additionally, investigating
the interactions of radial modes with the vertical
structures and instabilities, possible surface MRI, and even
gravitational instabilities on PPD substructure formation
could also yield valuable insights for future
research. Given that substructures in protoplanetary disks
are key to understanding the formation and evolution of
planetary systems, this line of research is vital in
unlocking some of the mysteries surrounding the early stages
of planetary formation.

\bigskip
\noindent
L. Wang and S. Xu acknowledge the computation resources
provided by the Kavli Institute of Astronomy and
Astrophysics at Peking University. We thank our collegues
for helpful discussions (alphabetical order of the last
names): Can Cui, Gregory Herczeg, Xiao Hu, Haifeng Yang,
Xinyu Zheng.

\bibliography{perturb}

\begin{thebibliography}{}
\expandafter\ifx\csname natexlab\endcsname\relax\def\natexlab#1{#1}\fi
\providecommand{\url}[1]{\href{#1}{#1}}

\bibitem[{{Bacciotti} {et~al.}(2000){Bacciotti}, {Mundt}, {Ray},
  {Eisl{\"o}ffel}, {Solf}, \& {Camezind}}]{2000ApJ...537L..49B}
{Bacciotti}, F., {Mundt}, R., {Ray}, T.~P., {et~al.} 2000, \apjl, 537, L49

\bibitem[{{Bai}(2013)}]{2013ApJ...772...96B}
{Bai}, X.-N. 2013, \apj, 772, 96

\bibitem[{{Bai}(2017)}]{2017ApJ...845...75B}
---. 2017, \apj, 845, 75

\bibitem[{{Bai} \& {Stone}(2011)}]{2011ApJ...736..144B}
{Bai}, X.-N., \& {Stone}, J.~M. 2011, \apj, 736, 144

\bibitem[{{Bai} \& {Stone}(2013)}]{2013ApJ...769...76B}
---. 2013, \apj, 769, 76

\bibitem[{{Bai} \& {Stone}(2017)}]{2017ApJ...836...46B}
---. 2017, \apj, 836, 46

\bibitem[{{Bai} {et~al.}(2016){Bai}, {Ye}, {Goodman}, \&
  {Yuan}}]{2016ApJ...818..152B}
{Bai}, X.-N., {Ye}, J., {Goodman}, J., \& {Yuan}, F. 2016, \apj, 818, 152

\bibitem[{{Balbus} \& {Hawley}(1991)}]{1991ApJ...376..214B}
{Balbus}, S.~A., \& {Hawley}, J.~F. 1991, \apj, 376, 214

\bibitem[{{Balbus} \& {Hawley}(1998)}]{1998RvMP...70....1B}
---. 1998, Reviews of Modern Physics, 70, 1

\bibitem[{{Banzatti} {et~al.}(2019){Banzatti}, {Pascucci}, {Edwards}, {Fang},
  {Gorti}, \& {Flock}}]{2019ApJ...870...76B}
{Banzatti}, A., {Pascucci}, I., {Edwards}, S., {et~al.} 2019, \apj, 870, 76

\bibitem[{{B{\'e}thune} {et~al.}(2017){B{\'e}thune}, {Lesur}, \&
  {Ferreira}}]{2017A&A...600A..75B}
{B{\'e}thune}, W., {Lesur}, G., \& {Ferreira}, J. 2017, \aap, 600, A75

\bibitem[{{Burrows} {et~al.}(1996){Burrows}, {Stapelfeldt}, {Watson}, {Krist},
  {Ballester}, {Clarke}, {Crisp}, {Gallagher}, {Griffiths}, {Hester},
  {Hoessel}, {Holtzman}, {Mould}, {Scowen}, {Trauger}, \&
  {Westphal}}]{1996ApJ...473..437B}
{Burrows}, C.~J., {Stapelfeldt}, K.~R., {Watson}, A.~M., {et~al.} 1996, \apj,
  473, 437

\bibitem[{{Chiang} \& {Goldreich}(1997)}]{1997ApJ...490..368C}
{Chiang}, E.~I., \& {Goldreich}, P. 1997, \apj, 490, 368

\bibitem[{{Fang} {et~al.}(2023{\natexlab{a}}){Fang}, {Pascucci}, {Edwards},
  {Gorti}, {Hillenbrand}, \& {Carpenter}}]{2023ApJ...945..112F}
{Fang}, M., {Pascucci}, I., {Edwards}, S., {et~al.} 2023{\natexlab{a}}, \apj,
  945, 112

\bibitem[{{Fang} {et~al.}(2018){Fang}, {Pascucci}, {Edwards}, {Gorti},
  {Banzatti}, {Flock}, {Hartigan}, {Herczeg}, \&
  {Dupree}}]{2018ApJ...868...28F}
---. 2018, \apj, 868, 28

\bibitem[{{Fang} {et~al.}(2023{\natexlab{b}}){Fang}, {Wang}, {Herczeg},
  {Hashimoto}, {Xu}, {Nemer}, {Pascucci}, {Haffert}, \&
  {Aoyama}}]{2023NatAs...7..905F}
{Fang}, M., {Wang}, L., {Herczeg}, G.~J., {et~al.} 2023{\natexlab{b}}, Nature
  Astronomy, 7, 905

\bibitem[{{Flores-Rivera} {et~al.}(2023){Flores-Rivera}, {Flock}, {Kurtovic},
  {Husemann}, {Banzatti}, {Ringqvist}, {Kamann}, {M{\"u}ller}, {Fendt},
  {Garc{\'\i}a Lopez}, {Marleau}, {Henning}, {Carrasco-Gonz{\'a}lez}, {van
  Boekel}, {Keppler}, {Launhardt}, \& {Aoyama}}]{2023A&A...670A.126F}
{Flores-Rivera}, L., {Flock}, M., {Kurtovic}, N.~T., {et~al.} 2023, \aap, 670,
  A126

\bibitem[{{Gressel} {et~al.}(2015){Gressel}, {Turner}, {Nelson}, \&
  {McNally}}]{2015ApJ...801...84G}
{Gressel}, O., {Turner}, N.~J., {Nelson}, R.~P., \& {McNally}, C.~P. 2015,
  \apj, 801, 84

\bibitem[{{Harrison} {et~al.}(2021){Harrison}, {Looney}, {Stephens}, {Li},
  {Teague}, {Crutcher}, {Yang}, {Cox}, {Fern{\'a}ndez-L{\'o}pez}, \&
  {Shinnaga}}]{2021ApJ...908..141H}
{Harrison}, R.~E., {Looney}, L.~W., {Stephens}, I.~W., {et~al.} 2021, \apj,
  908, 141

\bibitem[{{Hu} {et~al.}(2022){Hu}, {Li}, {Zhu}, \&
  {Yang}}]{2022MNRAS.516.2006H}
{Hu}, X., {Li}, Z.-Y., {Zhu}, Z., \& {Yang}, C.-C. 2022, \mnras, 516, 2006

\bibitem[{{Hu} {et~al.}(2019){Hu}, {Zhu}, {Okuzumi}, {Bai}, {Wang}, {Tomida},
  \& {Stone}}]{2019ApJ...885...36H}
{Hu}, X., {Zhu}, Z., {Okuzumi}, S., {et~al.} 2019, \apj, 885, 36

\bibitem[{{Ilgner} \& {Nelson}(2008)}]{2008A&A...483..815I}
{Ilgner}, M., \& {Nelson}, R.~P. 2008, \aap, 483, 815

\bibitem[{{Latter} {et~al.}(2010){Latter}, {Fromang}, \&
  {Gressel}}]{2010MNRAS.406..848L}
{Latter}, H.~N., {Fromang}, S., \& {Gressel}, O. 2010, \mnras, 406, 848

\bibitem[{{Lavalley-Fouquet} {et~al.}(2000){Lavalley-Fouquet}, {Cabrit}, \&
  {Dougados}}]{2000A&A...356L..41L}
{Lavalley-Fouquet}, C., {Cabrit}, S., \& {Dougados}, C. 2000, \aap, 356, L41

\bibitem[{{Lesur} {et~al.}(2014){Lesur}, {Kunz}, \&
  {Fromang}}]{2014A&A...566A..56L}
{Lesur}, G., {Kunz}, M.~W., \& {Fromang}, S. 2014, \aap, 566, A56

\bibitem[{{Leung} \& {Ogilvie}(2020)}]{2020MNRAS.498..750L}
{Leung}, P. K.~C., \& {Ogilvie}, G.~I. 2020, \mnras, 498, 750

\bibitem[{{McGinnis} {et~al.}(2018){McGinnis}, {Dougados}, {Alencar},
  {Bouvier}, \& {Cabrit}}]{2018AA...620A..87M}
{McGinnis}, P., {Dougados}, C., {Alencar}, S.~H.~P., {Bouvier}, J., \&
  {Cabrit}, S. 2018, \aap, 620, A87

\bibitem[{{McNally} \& {Pessah}(2015)}]{2015ApJ...811..121M}
{McNally}, C.~P., \& {Pessah}, M.~E. 2015, \apj, 811, 121

\bibitem[{{Nemer} {et~al.}(2020){Nemer}, {Goodman}, \&
  {Wang}}]{2020ApJ...904L..27N}
{Nemer}, A., {Goodman}, J., \& {Wang}, L. 2020, \apjl, 904, L27

\bibitem[{{Noriega-Crespo} {et~al.}(2002){Noriega-Crespo}, {Cotera}, {Young},
  \& {Chen}}]{2002ApJ...580..959N}
{Noriega-Crespo}, A., {Cotera}, A., {Young}, E., \& {Chen}, H. 2002, \apj, 580,
  959

\bibitem[{{Owen} {et~al.}(2012){Owen}, {Clarke}, \&
  {Ercolano}}]{2012MNRAS.422.1880O}
{Owen}, J.~E., {Clarke}, C.~J., \& {Ercolano}, B. 2012, \mnras, 422, 1880

\bibitem[{{Pascucci} {et~al.}(2023){Pascucci}, {Cabrit}, {Edwards}, {Gorti},
  {Gressel}, \& {Suzuki}}]{2023ASPC..534..567P}
{Pascucci}, I., {Cabrit}, S., {Edwards}, S., {et~al.} 2023, in Astronomical
  Society of the Pacific Conference Series, Vol. 534, Protostars and Planets
  VII, ed. S.~{Inutsuka}, Y.~{Aikawa}, T.~{Muto}, K.~{Tomida}, \& M.~{Tamura},
  567

\bibitem[{{Rigliaco} {et~al.}(2013){Rigliaco}, {Pascucci}, {Gorti}, {Edwards},
  \& {Hollenbach}}]{2013ApJ...772...60R}
{Rigliaco}, E., {Pascucci}, I., {Gorti}, U., {Edwards}, S., \& {Hollenbach}, D.
  2013, \apj, 772, 60

\bibitem[{{Riols} {et~al.}(2020){Riols}, {Lesur}, \&
  {Menard}}]{2020A&A...639A..95R}
{Riols}, A., {Lesur}, G., \& {Menard}, F. 2020, \aap, 639, A95

\bibitem[{{Sano} \& {Miyama}(1999)}]{1999ApJ...515..776S}
{Sano}, T., \& {Miyama}, S.~M. 1999, \apj, 515, 776

\bibitem[{{Sarafidou} {et~al.}(2024){Sarafidou}, {Gressel}, {Picogna}, \&
  {Ercolano}}]{2024MNRAS.530.5131S}
{Sarafidou}, E., {Gressel}, O., {Picogna}, G., \& {Ercolano}, B. 2024, \mnras,
  530, 5131

\bibitem[{{Simon} {et~al.}(2013{\natexlab{a}}){Simon}, {Bai}, {Armitage},
  {Stone}, \& {Beckwith}}]{2013ApJ...775...73S}
{Simon}, J.~B., {Bai}, X.-N., {Armitage}, P.~J., {Stone}, J.~M., \& {Beckwith},
  K. 2013{\natexlab{a}}, \apj, 775, 73

\bibitem[{{Simon} {et~al.}(2013{\natexlab{b}}){Simon}, {Bai}, {Stone},
  {Armitage}, \& {Beckwith}}]{2013ApJ...764...66S}
{Simon}, J.~B., {Bai}, X.-N., {Stone}, J.~M., {Armitage}, P.~J., \& {Beckwith},
  K. 2013{\natexlab{b}}, \apj, 764, 66

\bibitem[{{Simon} {et~al.}(2016){Simon}, {Pascucci}, {Edwards}, {Feng},
  {Gorti}, {Hollenbach}, {Rigliaco}, \& {Keane}}]{2016ApJ...831..169S}
{Simon}, M.~N., {Pascucci}, I., {Edwards}, S., {et~al.} 2016, \apj, 831, 169

\bibitem[{{Stapelfeldt} {et~al.}(1999){Stapelfeldt}, {Watson}, {Krist},
  {Burrows}, {Crisp}, {Ballester}, {Clarke}, {Evans}, {Gallagher}, {Griffiths},
  {Hester}, {Hoessel}, {Holtzman}, {Mould}, {Scowen}, \&
  {Trauger}}]{1999ApJ...516L..95S}
{Stapelfeldt}, K.~R., {Watson}, A.~M., {Krist}, J.~E., {et~al.} 1999, \apjl,
  516, L95

\bibitem[{{Turner} {et~al.}(2014){Turner}, {Fromang}, {Gammie}, {Klahr},
  {Lesur}, {Wardle}, \& {Bai}}]{2014prpl.conf..411T}
{Turner}, N.~J., {Fromang}, S., {Gammie}, C., {et~al.} 2014, Protostars and
  Planets VI, 411

\bibitem[{{Turner} {et~al.}(2007){Turner}, {Sano}, \&
  {Dziourkevitch}}]{2007ApJ...659..729T}
{Turner}, N.~J., {Sano}, T., \& {Dziourkevitch}, N. 2007, \apj, 659, 729

\bibitem[{{Vlemmings} {et~al.}(2019){Vlemmings}, {Lankhaar}, {Cazzoletti},
  {Ceccobello}, {Dall'Olio}, {van Dishoeck}, {Facchini}, {Humphreys},
  {Persson}, {Testi}, \& {Williams}}]{2019A&A...624L...7V}
{Vlemmings}, W.~H.~T., {Lankhaar}, B., {Cazzoletti}, P., {et~al.} 2019, \aap,
  624, L7

\bibitem[{{Wang} {et~al.}(2019){Wang}, {Bai}, \&
  {Goodman}}]{2019ApJ...874...90W}
{Wang}, L., {Bai}, X.-N., \& {Goodman}, J. 2019, \apj, 874, 90

\bibitem[{{Wang} \& {Goodman}(2017{\natexlab{a}})}]{2017ApJ...847...11W}
{Wang}, L., \& {Goodman}, J. 2017{\natexlab{a}}, \apj, 847, 11

\bibitem[{{Wang} \& {Goodman}(2017{\natexlab{b}})}]{2017ApJ...835...59W}
---. 2017{\natexlab{b}}, \apj, 835, 59

\bibitem[{{Wardle}(2007)}]{2007Ap&SS.311...35W}
{Wardle}, M. 2007, \apss, 311, 35

\bibitem[{{Wardle} \& {K\"onigl}(1993)}]{Wardle+Konigl1993}
{Wardle}, M., \& {K\"onigl}, A. 1993, \apj, 410, 218

\bibitem[{{Watson} \& {Stapelfeldt}(2004)}]{2004ApJ...602..860W}
{Watson}, A.~M., \& {Stapelfeldt}, K.~R. 2004, \apj, 602, 860

\bibitem[{{Woitas} {et~al.}(2002){Woitas}, {Ray}, {Bacciotti}, {Davis}, \&
  {Eisl{\"o}ffel}}]{2002ApJ...580..336W}
{Woitas}, J., {Ray}, T.~P., {Bacciotti}, F., {Davis}, C.~J., \&
  {Eisl{\"o}ffel}, J. 2002, \apj, 580, 336

\bibitem[{{Xu} \& {Bai}(2016)}]{2016ApJ...819...68X}
{Xu}, R., \& {Bai}, X.-N. 2016, \apj, 819, 68

\end{thebibliography}
\bibliographystyle{aasjournal}

\appendix
\nopagebreak

\section{Extra terms in the steady-state solution}
\label{sec:apdx-xtra-terms}

The general guideline in simplifying the equations for the
local wind-driven accretion model is to ignore the terms
with $R$ on their denominators or with derivatives of
$R$. In \S\ref{sec:method-steady} the inclusion of
$\partial_R(v_\varphi B_R)$ (as part of the $\partial_RE_z$
term) has been explained, and this appendix section
discusses two other terms with $\mu_k$ on the prospectively
important denominators.

One of them is associated with $J_z$, which can be expressed
as the derivatives of magnetic fields,
\begin{equation}
  \label{eq:apdx-xtra-terms-jz-dim}
  \dfrac{4\pi J_z}{c} = \dfrac{\partial_R (R B_\varphi) -
    \partial_\varphi B_R}{R} = 
  \dfrac{B_\varphi}{R} + \partial_R B_\varphi .
\end{equation}
Note that $\partial_\varphi = 0$ strictly holds due to the
axisymmetry.  While it is unable to examine whether
$\partial_R B_\varphi$ plays an essential role with radially
local models, the rest ($B_\varphi/R$) could still make the
terms involving $J_z B_\varphi$ important with sufficiently
large $|B_\varphi|$. In the dimensionless form, this product
reads $j_zb_\varphi \simeq b_\varphi^2/\mu_k$. As one can
observe from the numerical and semi-analytic results in
\S\ref{sec:fid-steady} and \S\ref{sec:var-models},
$|b_\varphi|$ could reach $\sim 10^0-10^1$, making the
$j_zb_\varphi$ term important with $\mu_k\sim 30$
(eq.~\ref{eq:method-muk}) or even smaller at larger disk
radii. Therefore, although $j_z\sim 0$ in eq.~\eqref
{eq:method-system-efield} essentially holds in most cases,
we still include the $j_z$ term when the steady-state
solutions are compared to the simulation results in
\S\ref{sec:fid-steady}.

The other term is related to the gravitational force. At
$z=0$, a local model assuming negligible radial pressure
gradient has $\partial_R \Phi = v_k^2/R$, where $v_k$ is the
Keplerian velocity. In the derivation of the radial momentum
equation in eq.~\eqref{eq:method-system-ode}, this equality
is assumed to be true at all altitudes. However, one can
prove that $|\partial_R \Phi - v_k^2/R|$ may be no longer
negligible at sufficiently high $z$.
This leads to an extra term $g_R$ in the radial momentum
equation (eq.~\ref{eq:method-system-ode}), using
$z / R = \zeta / \mu_k$, and
$\partial_z \rightarrow (\mu_k/R) \partial_\zeta$,
\begin{equation}
  \label{eq:apdx-xtra-terms-phir}
  g_R \equiv - \dfrac{R(\partial_R\Phi - v_k^2/R)}
  {\mu_k c_s^2} = \dfrac{v_k^2}{\mu_k c_s^2}
  \left[ 1 - \left( 1+ \dfrac{z^2}{R^2} \right)^{-3/2} 
  \right] = \mu_k
  \left[ 1 - \left( 1+ \dfrac{\zeta^2}{\mu_k^2} \right)^{-3/2}
  \right]\ .
\end{equation}
This approach of approximating radial gravity force is the
same as the vertically global shearing box methods
elaborated in \citet{2015ApJ...811..121M}. When we consider
the disk surface, $\zeta\sim 3-4$, and the $g_R$ could also
reach the order of unity. Similar to the terms related to
$j_z$, the $g_R$ is also included to reach better agreements
with numerical simulation results. One may notice that a
similar correction can also be applied to the vertical
momentum equation, yet the other terms dwarf the value of
the resulting correction.

\section{Electric current equations for steady-state solutions}
\label{sec:apdx-current}

Assuming $\mu_z\rightarrow 0$, eqs.~\eqref
{eq:method-system-ode} yield the dimensionless velocity
components $\mu_{R,\varphi}$,
\begin{equation}
  \label{eq:apdx-steady-vel}
  \mu_{\varphi} = \dfrac{j_zb_\phi - j_\varphi}
  {\beta_0\varrho}  - \dfrac{g_R}{2} ,
  \quad
  \mu_R = \frac{4 (j_zb_R - j_R)} {\beta_0\varrho}.
\end{equation}
Combining these $\mu_{R,\varphi}$ expressions with
eqs.~\eqref {eq:method-system-efield} and \eqref
{eq:method-system-nonideal}, one gets the equations which
shall be solved for $j_{R,\varphi}$,
\begin{equation}
  \label{eq:method-ode-aux}
  \begin{bmatrix}
    \alpha_{R R} & \alpha_{R \varphi} + 1/(\beta_0\varrho)\\
    \alpha_{\varphi R} - 4/(\beta_0\varrho)
                 & \alpha_{\varphi \varphi}
  \end{bmatrix}
  \begin{bmatrix} j_R\\ j_{\varphi} \end{bmatrix} =
  \begin{bmatrix}
    \tilde{\varepsilon}_R - g_R / 2
    + j_z b_\varphi/(\beta_0\varrho) \\
    -4j_zb_R/(\beta_0\varrho)
  \end{bmatrix},
\end{equation}
where $\tilde{\varepsilon}_R\equiv \varepsilon_R + \mu_k$ is
the dimensionless electric field in the Keplerian rotating
frame.

\section{Matching the ideal MHD wind solutions}
\label{sec:apdx-matching-wind}

\citet{2016ApJ...818..152B} introduced a method of
constructing magnetized isothermal wind models above disk
surfaces, which is briefly summarized here. This method is
based on the conserved quantities along the magnetic field
lines, including the Bernoulli parameter $H$ (following from
the energy conservation), the mass-field flux ratio $k$
(following from the continuity equation; not to be confused
with dimensionless wavenumbers in \S\ref{sec:simp-modes}),
and the field line angular velocity parameter $\omega$,
reading,
\begin{equation}
  \label{eq:apdx-ideal-wind-pars}
  H  = \dfrac{B_\p^2}{2k^2x^2} + \dfrac{\omega^2R^2}{2} 
  \left[ \left( \dfrac {R_\A^2/R^2-1} {x-1} \right)^2 -
    1 \right] + h - \dfrac{G M_*}{(R^2+z^2)^{1/2}}\ ;\quad
  k  \equiv \dfrac{4\pi\rho v_\p}{B_\p}  ,\quad
  \omega = \Omega - \dfrac{k B_\varphi} {4\pi \rho R} .
\end{equation}
Here $M_*$ is the stellar mass, $x\equiv 4\pi \rho / k^2$ is
the density parameter, the subscripts ``p'' in $v_\p$ and
$B_\p$ indicate the poloidal components of velocities and
magnetic fields, $R$ and $z$ are the cylindrical coordinates
of the spatial point on the current field line, and $R_\A$
marks the radius of the Alfv{\' e}nic point on the same
field line. The thermodynamics of gas elements are
  involved by the specific enthalpy,
\begin{equation}
  \label{eq:apdx-ideal-wind-enthalpy}
  h = \int \dfrac{\d p}{\rho} =
  \begin{cases}
    & c_s^2\ln \left(\dfrac{\rho}{\rho_0}\right) \ ,\quad
      \gamma = 1\  \text{(Isothermal)}\ ;
    \\
    & c_{s0}^2\left(\dfrac{\gamma}{\gamma - 1}\right)
      \left(\dfrac{\rho}{\rho_0}\right)^{\gamma - 1},
      \quad \gamma > 1 \ ,
  \end{cases}
\end{equation}
where $\rho_0$ is the wind-base mass density (and the
subscript ``0'' in general marks the quantities at the wind
base in this appendix. In our numerical experiments, we
found that the MHD profiles yielded by prescribing
$\gamma = 7/5$ fit the results in axisymmetric global
simulations better than other choices (note that this
$\gamma$ index only applies to the wind, and the disk below
the wind base is still vertically isothermal). Since the
heating processes largely stop after wind gases finish
acceleration and join the wind, this choice also stands
closer to a MHD dominated wind whose major component is
molecular due to self-shielding and cross-shielding effects
of photodissociation \citep[e.g.][]{2019ApJ...874...90W}. 
The poloidal field $B_\p$ in the wind is prescribed as,
\begin{equation}
  \label{eq:apdx-ideal-wind-bp}
  B_\p = B_{\p 0} \dfrac{1+q}{(R/R_0) + q(R/R_0)^2} ,
\end{equation}
in which the parameter
$q$ controls the poloidal fields' transition from parallel
to diverging. In practice, we find that $q =
1$ can yield accretion and wind solutions that match the
simulation results best. The vertical coordinate
$z$ is related to
$R$ along the designated field line by a straight-line
geometry, $z = z_0 + (R-R_0) \tan
\theta$. 
We adopt $\theta =
\tan^{-1}(b_z/b_R)$ at the wind base to guarantee the
continuity of magnetic fields. The configuration of poloidal
fields in the wind can be calculated accordingly. Physically
plausible solutions are obtained by,
\begin{equation}
  \label{eq:apdx-ideal-wind-sol}
  H = E ,\quad \dfrac{\partial H}{\partial x} = 0 ,\quad
  \dfrac{\partial H}{\partial R} = 0 .
\end{equation}

In order to construct consistent wind-driven accretion
models, the accretion solution described in
\S\ref{sec:non-ideal-mhd} has to be connected to the
solution described by eq.~\eqref {eq:apdx-ideal-wind-pars}
at the wind base, by matching key physical quantities. We
first convert eqs.~\eqref{eq:apdx-ideal-wind-pars},
\eqref{eq:apdx-ideal-wind-bp} into the dimensionless form,
\begin{equation}
  \label{eq:apdx-ideal-wind-dimless}
  \begin{split}
    & \mathcal{H} \equiv \dfrac{H}{c_s^2} = \left(
      \dfrac{b_{\p 0}^2}{2\tilde{k}^2x^2}  \right)
      \left( \dfrac{1+q} {\tilde{R}^2 + q\tilde{R}^2}
      \right)   + \dfrac{\tilde{\omega}^2\tilde{R}^2}{2}
      \left[ \left( \dfrac{\tilde{R}_\A^2/\tilde{R}^2-1}
      {x-1} \right)^2 - 1 \right]
      + \left(\dfrac{\gamma}{\gamma - 1}\right)
      \left(\dfrac{x}{x_{\rm wb}}\right)^{\gamma - 1}
      -\dfrac{\mu_k^2}{\tilde{r}(R)}\ ;
    \\
    & b_{\p 0} \equiv \dfrac{B_{\p 0}}{B_{z0}}\ ,\quad
      \tilde{R} \equiv \dfrac{R}{R_0}\ ,\quad
      \tilde{R}_\A \equiv \dfrac{R_\A}{R_0}\ ;\
      x_{\rm wb} = \left[\left( \dfrac{2} {\beta_0\varrho} 
      \right)\tilde{k} \right]_{\rm wb}^{-1},\ 
      \tilde{\omega} \equiv \dfrac{\omega R_0}{c_s}
      = \left[(\mu_\varphi + \mu_k) - \left(
      \dfrac{2}{\beta_0\varrho}\right)\tilde{k}b_\phi
      \right]_{\rm wb}, \\
  \end{split}
\end{equation}
where the subscript ``wb'' indicate the wind-base values.
The dimensionless $\tilde{k} \equiv k c_s /
B_{z0}$ can be obtained by solving the following equation
deduced from other conservation quantities,
\begin{equation}
  \label{eq:apdx-ideal-wind-k}
  \left(\dfrac{2 b_{\varphi, {\rm wb}}\tilde{R}_\A^2}
    {\beta_0 \varrho_{\rm wb}} \right) \tilde{k}^2 -
  (\mu_k+\mu_{\varphi, {\rm wb}}) (\tilde{R}_\A^2 - 1)
  \tilde{k} - b_{\varphi, {\rm wb}} = 0 \ .
\end{equation}
Each value of
$\tilde{k}$ indicates a mass load condition of the wind,
which corresponds to a plausibale
$\tilde{R}_\A$ value that allows the wind solution to get
through the slow and fast magneto-sonic points smoothly.


We define the wind base at $\mu_R = 0$, separating the
accreting and wind-launching regions.  The following
procedures are taken to obtain a set of solutions with
matched wind and accretion profiles:
\begin{enumerate}
\item Select a $\varepsilon_{R0}$ (the radial component of
  the mid-plane dimensionless electric field; see
  \S\ref{sec:method-steady}) and integrate the steady-state
  solution to the wind base;
\item Calculate the values of $b_{\p 0}$,
  $\tilde{\omega}$,
  and $\mathcal{H}_{\rm wb}$ at the wind base;
\item Choose an
  $\tilde{R}_\A$ value and solve for a physically plausible
  wind solution
  by solving the algebraic equations
  $\partial \mathcal{H}/\partial x = \partial
  \mathcal{H}/\partial \tilde{R} = 0$ for the regularity
  conditions at both slow and fast magnetosonic points (note
  that the value of $\tilde{k}$ is obtained by solving
  eq.~\ref{eq:apdx-ideal-wind-k});
\item Adjust $\varepsilon_{R0}$ and conduct Steps 1 through
  3 iteratively, so that the $\mathcal{H}$ value in Step 3
  equals to the $\mathcal{H}_{\rm wb}$ in Step 2.
\end{enumerate}
We have noticed that matching a wind solution to a radially
local accretion solution inevitably leads to the
inconsistency of mass conservation, either at the matching
point (e.g., this work; \citealt{2016ApJ...818..152B}) or at
the equatorial plane
\citep[e.g.][]{Wardle+Konigl1993}. However, such
inconsistency does not undermine the validity of the
background MHD profiles, on which subsequent studies (e.g.,
the perturbations) are carried out.

\section{Boundary Conditions of Perturbation Modes}

\label{sec:apdx-boundary}

Inside the wind, the perturbations should be evaluated along
the magnetic field lines, and the perturbed quantities
should involve $\mu_z$ and $\partial_R $.  The radial
derivative operator along a field line in the wind satisfies
$\partial_R \rightarrow \tan\theta\ \partial_z$.  Still
assuming axisymmetry ($\partial_\varphi \rightarrow 0$), the
perturbed induction equations read,
\begin{equation}
  \label{eq:apdx-bnd-wind-induction}
  \nu \delta b_\varphi = -\partial_z \delta
  \varepsilon_R
  + \tan\theta\,\partial_z\varepsilon_z \ ,\ 
  \nu \delta b_R = \partial_\zeta
  \delta \varepsilon_\varphi\ .
\end{equation}
As this work aims at the radially local modes that
intentionally excludes radial transport of magnetic fluxes,
the perturbation $\delta \varepsilon_\varphi$ should vanish,
which directly leads to $\delta b_R = 0$.  Meanwhile, the
perturbed toroidal electric field in the fluid frame
$\delta \varepsilon'_{\varphi}$ should vanish under high
conductivity, and the relation
$ \delta \varepsilon'_{\varphi} = \delta
\varepsilon_{\varphi} + b_R \delta \mu_z + \mu_z \delta b_R
- b_z \delta \mu_R - \mu_R \delta b_z$ also leads to
$\delta b_R = 0$ by assuming that the perturbations follow
the field line ($\delta b_z = \tan\theta\,\delta b_R$).
This constraint indeed naturally guarantees the solenoidal
condition of the perturbed fields, which reads
$\nabla \cdot \delta \B = \delta B_R / R + \partial_R \delta
B_R + \partial_z \delta B_z = 0$.

With $\delta b_R = 0$ in mind, the perturbation equations
inside the wind then read
$C_1 \partial_\zeta \delta \tilde{b} = C_0 \delta
\tilde{b}$, in which
$\delta \tilde{b}\equiv [\delta b_{\varphi}, \delta \mu_R,
\delta \mu_\phi]^T$ and,
\begin{equation}
  \label{eq:apdx-bnd-wind-matrix}
   C_1 \equiv 2 \tan\theta\times
    \begin{bmatrix}
      \mu_R & b_\varphi & - b_R \\
      2 b_\varphi / (\beta_0 \varrho ) & \mu_R & 0 \\
      -2 b_\varphi / (\beta_0 \varrho ) & 0 & 2 \mu_R 
    \end{bmatrix}\ ,\quad
     C_0 \equiv \nu I + 2 \tan\theta
     \ \times
    \begin{bmatrix}
      \partial_\zeta\mu_R & \partial_\zeta b_\varphi
      & - \partial_\zeta b_R \\
      \partial_\zeta b_\varphi / (\beta_0 \varrho )
      & \partial_\zeta \mu_R & -\cot\theta \\
      0 & (\cot\theta)/4 + \partial_\zeta \mu_\varphi & 0 
    \end{bmatrix}\ .
\end{equation}

At the wind base $\zeta_{\rm wb}$, a perturbation mode is
connected to the wind perturbations
(eqs.~\ref{eq:apdx-bnd-wind-matrix}) by adopting the
$\delta\mu_R$ and $\delta \mu_\phi$ obtained from
eqs.~\eqref{eq:method-perturb-deriv-mom}, and the
integration continues to an stopping altitude
$\zeta_1 \geq \zeta_{\rm wb}$. Ideally, one could set
$\zeta_1\rightarrow \infty$ and aim at vanishing
$\{\delta b_R, \delta b_\varphi\}$ by adjusting the
eigenvalue $\nu$ and the free parameter
$\delta j_{\varphi 0}$ (symmetric) or $\delta b_{\varphi 0}$
(anti-symmetric).  However, the integration to infinity is
difficult and unnecessary, especially when one considers the
difficulties of perturbation modes passing through the
critical points (including the poloidal Alfv{\' e}nic point,
and the slow and fast mangetosonic points). In a variety of
numerical experiments, we choose different finite $\zeta_1$
values, to verify that $\nu$ changes by no more than
$\sim 10\%$ as long as $\zeta_1 \geq \zeta_{\rm wb}$, even
if $\zeta_1$ is located at the poloidal Alfv{\' e}nic point
(Appendix~\ref{sec:apdx-matching-wind}) that is typically
$\gtrsim 10$ scale heights above $\zeta_{\rm wb}$. This can
also be qualitatively explained if one inspects
eq.~\eqref{eq:apdx-bnd-wind-matrix} for an leading-order
asymptotic approximation regarding $\varrho^{-1}$ and
$\delta b_\varphi$,
\begin{equation}
  b_\varphi \partial_\zeta \delta b_\varphi \simeq
  \delta b_\varphi \partial_\zeta b_\varphi \ \Rightarrow\
  \delta b_\varphi \proptosim b_\varphi\ ,
\end{equation}
which holds when $\varrho \ll 1$, a quite typical condition
within the wind region. Since $B_\varphi\propto R^{-1}$
inside the magneto-thermal wind
\citep[e.g.][]{2016ApJ...818..152B}, $\delta b_\varphi$
decays rather slowly at relatively large
altitudes. Therefore, the altitude of setting
$\delta b_\varphi = 0$ does {\it not} affect the eigenmodes
significantly. For simplicity and clarity, we fix
$\zeta_1 = \zeta_{\rm wb}$ and adjust the free parameters so
that
$\delta b_R |_{\zeta = \zeta_1} = \delta b_{\varphi}
|_{\zeta = \zeta_1} = 0$.


\end{document}